\documentclass[12pt]{elsarticle}
\usepackage{amsmath,amssymb,amsthm,amsfonts}
\usepackage{caption}
\usepackage[title]{appendix}
\setcitestyle{authoryear,open={},close={}}

\usepackage{graphicx,epsfig}
\usepackage{psfrag}
\usepackage[usenames]{color}
\usepackage[hmargin=1in,vmargin=1in]{geometry}
\usepackage{stabular,morefloats}
\usepackage{textcomp}
\usepackage{titlesec}
\usepackage{colortbl}
\usepackage{xcolor}
\usepackage{multirow}
\usepackage{mathrsfs}
\usepackage{tabularx}
\usepackage{comment}
\usepackage{multirow,tabularx,afterpage,fancyhdr}
\usepackage{placeins}
\titleformat{\subsubsection}
  {\normalfont\selectfont}{\thesubsubsection}{1em}{}
\numberwithin{equation}{section}
\usepackage{float}
\usepackage{xcolor}
\usepackage{soul}
\definecolor{lightblue}{rgb}{0,0,0.1}
\sethlcolor{lightblue}

\usepackage[subrefformat=parens,labelformat=parens]{subfig}

\colorlet{Changes@Color}{black}
\makeatletter
\renewcommand{\@biblabel}[1]{[#1]\hfill}
\makeatother

\usepackage{etoolbox}
\makeatletter
\patchcmd{\ps@pprintTitle}
  {Elsevier}
  {the International Journal of Plasticity}
  {}{}
\makeatother

\begin{document}

\title{A large deformation model for quasi-static to high strain rate response of a rate-stiffening soft polymer}
\author[main]{Aditya Konale}
\author[main,main2]{Zahra Ahmed}
\author[main3]{Piyush Wanchoo}
\author[main,main2]{Vikas Srivastava\corref{cor1}}
\ead{vikas\_srivastava@brown.edu}
\cortext[cor1]{Corresponding author}
\address[main]{School of Engineering, Brown University, Providence, RI 02912, USA}
\address[main2]{Center for Biomedical Engineering, Brown University, Providence, RI 02912, USA}
\address[main3]{Department of Mechanical, Industrial and Systems Engineering, University of Rhode Island, Kingston, RI, 02881, USA}

\begin{abstract}
Polyborosiloxane (PBS) is an important rate-stiffening soft polymer with dynamic, reversible crosslinks used in applications ranging from self-healing sensing and actuation to body and structural protection. Its highly rate-dependent response, especially for impact-mitigating structures, is important. However, the large strain response of PBS has not been characterized over quasi-static to high strain rates. Currently, there are no constitutive models that can predict the strongly rate-dependent large-deformation elastic-viscoplastic response of PBS. To address this gap, we have developed a microstructural physics motivated constitutive model for PBS, similar soft polymers and polymer gels with dynamic crosslinks to predict their large strain, non-linear loading-unloading, and significantly rate-dependent response. We have conducted compression experiments on PBS up to true strains of $\sim$125$\%$ and over a wide strain rate range of 10$^{-3}$ s$^{-1}$ to 10$^{3}$ s$^{-1}$. The proposed semi-physical model fairly accurately captures the response of PBS over six decades of strain rates. We propose boron-oxygen coordinate-bond dynamic crosslinks with macroscopic relaxation timescale $\tau \approx 3$ s and temporary entanglement lockups at high strain rates acting as crosslinks with $\tau \approx 0.0005$ s as the two types of crosslink network mechanisms in PBS. We have outlined a numerical update procedure to evaluate the convolution-like time integrals arising from dynamic crosslink kinetics. Experiments involving three-dimensional inhomogeneous deformations were used to verify the predictive capabilities of our model and its finite element implementation. The modeling framework can be adopted for other dynamically crosslinked rate-stiffening soft polymers and polymer gels that are microstructurally similar to PBS.  
\end{abstract}

\begin{keyword}
Polymers; Rate-stiffening; Polyborosiloxane; Dynamic and reverisble crosslinks; Gels; Impact; High strain rate; Split Hopkinson Pressure Bar; Constitutive model
\end{keyword}

\maketitle
\newpage 
\section{Introduction}\label{sec1}

Soft polymers and polymer gels\footnote{Polymer gels are solvent swelled polymer networks (\citep{Chestergeldiffusion1}, \citep{Ekuhl3}, \citep{shukla2020}, \citep{Chestergeldiffusion2}). In this paper, we only consider and refer to polymer gels that have dynamic, reversible crosslinks and have negligible solvent diffusion in the mechanical loading timescales of interest, i.e., the concentration of the solvents is approximately constant both spatially and temporally. As an example, diffusion of solvent in a polymer gel can be neglected to analyze its mechanical response during a high rate loading event.} with dynamic, reversible crosslinks that exhibit rate-dependent mechanical behavior are an important class of materials with applications ranging from targeted drug delivery ({\citep{XZhaodrugdelivery, DrugDelivery, Balakrishnan, Toughhydrogelsreview}}), biological and structural adhesives (\citep{Yuk2016, Hong, XZhaoadhesive, XZhaotoughhydrogels2, XZhaoadhesive2}), mechanical sensing and actuation (\citep{Yuk2017, Liangsensors, Organmotionsensing, Bodymotionsensing, Bodymotionsensing2}) to flexible impact mitigators (\citep{Wangetal(2016)SSGPU,Liuetal(2020)foamB:Ocrosslinkref,Sangetal(2021)foamB:Ocrosslinkref}, \citep{Zongetal(2021)rheometryref,Tangetal(2022)B:Ocrosslinkrubberyref}). 

Polyborosiloxane (PBS), a key rate-stiffening soft polymer has dynamic boron (B)-oxygen (O) coordinate-bond crosslinks and exhibits significantly different mechanical behavior at different timescales (\citep{Lietal(2014)B:Ocrosslinkref, Wangetal(2014)B:Ocrosslinkref, Wangetal(2016)MagneticSSGSHPB, Tangetal(Synthesis), Wangetal(2018)AnovelmagneticSSGwithselfhealingability, Zhaoetal(2018)IHE, Liuetal(2019)B:Ocrosslinkref, ZHAOSSGReview, Kurkinetal(Synthesis), Liuetal(2022)B:Ocrosslinkref, Kim2022SSG}). PBS has been used to develop self-healing sensing and actuation interfaces (\citep{PBSsensingactuators}). It has been shown to mitigate impact loads (\citep{Zhangetal(2020)SSGMechanoluminescence, ZHAOSSGReview, Kurkinetal(Synthesis), Kim2022SSG}). Lightweight and flexible impact-mitigating composite structures where PBS is integrated into polymer foams (\citep{TU2023116811}) or sandwich structures (\citep{Myronidisetal(2022)rheometryref}) have been shown to provide protection against impact.  Material response characterization studies on PBS have been limited to small strain loadings, primarily oscillatory shear using a rheometer to quantify the effects of loading rates ({\citep{Wangetal(2014)B:Ocrosslinkref,Tangetal(Synthesis),Wangetal(2016)SSGPU,Wangetal(2018)AnovelmagneticSSGwithselfhealingability,Zhangetal(2020)SSGMechanoluminescence,Zhaoetal(2018)IHE,Kurkinetal(Synthesis),Zongetal(2021)rheometryref,Myronidisetal(2022)rheometryref}}). Large strain material response characterization studies have been conducted only on PBS with fillers to enhance its shape stability (\citep{Jiangetal(2014)IHPC,Zhaoetal(2018)IHE}) or to provide sensitivity to magnetic fields (\citep{Wangetal(2014)B:Ocrosslinkref, Wangetal(2016)MagneticSSGSHPB, Wangetal(2018)AnovelmagneticSSGwithselfhealingability,Liuetal(2019)B:Ocrosslinkref}).  It is important to study and model the response of pure PBS without any fillers to understand its physical material behavior. 

Despite the broad applications of PBS, its large strain material response has not been characterized over a wide range of strain rates. There is no large deformation constitutive model which can predict the response of PBS spanning the quasi-static (10$^{-3}$ s$^{-1}$) to high (10$^{3}$ s$^{-1}$) strain rate range. The time dependent behavior of pure PBS is shown through our experimental observations considering two distinct timescales:  (i) PBS flows under its weight over hours as shown in Figure \ref{Flow of PBS under its own weight, SHPB rubbery response}(a) where a PBS sphere flattens in an hour and has negligible shape recovery, and (ii) when subjected to a high loading rate, PBS exhibits notable shape recovery after undergoing significant deformation as seen in Figure \ref{Flow of PBS under its own weight, SHPB rubbery response}(b) where PBS is subjected to a strain rate of approximately 5000 s$^{-1}$ using a Split Hopkinson Pressure Bar (SHPB). Figure \ref{Flow of PBS under its own weight, SHPB rubbery response}(b) shows the PBS deformation over milliseconds. 

\begin{figure}[h!]
    \begin{center}
		\includegraphics[width=13 cm]{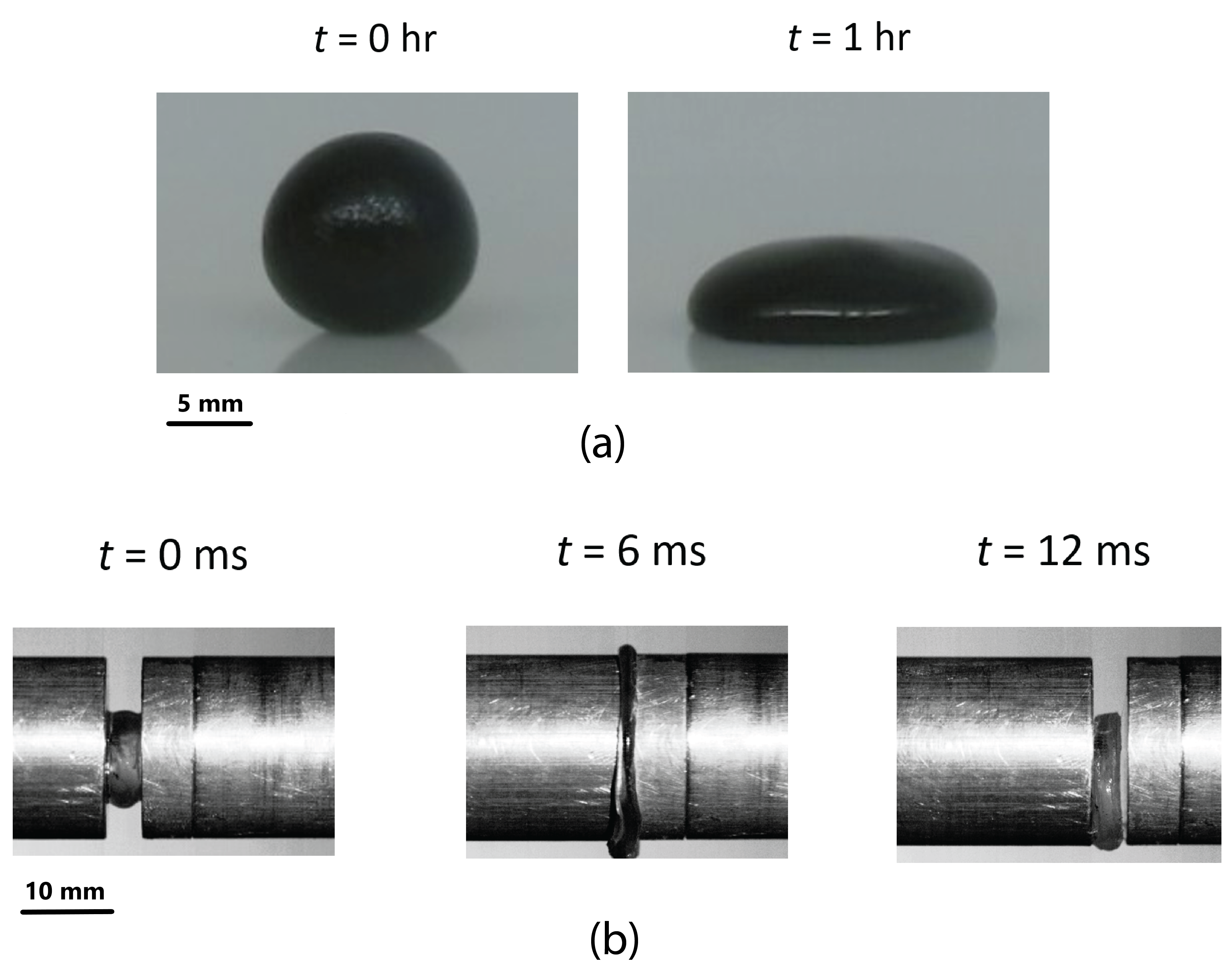}
	\end{center}
    \caption{{\small(a) Flow of a PBS sphere under its weight in 1 hour. The images are shown at $t$ = 0 and 1 hour. The PBS sphere was colored for easy visualization. (b) Response of PBS subjected to a strain rate loading of approximately 5000 s$^{-1}$ using a Split Hopkinson Pressure Bar. The PBS deformation images are shown at $t$ = 0, 6, and 12 milliseconds (ms). PBS experiences large deformation followed by significant shape recovery at high rates.}}
    \label{Flow of PBS under its own weight, SHPB rubbery response}
\end{figure} 

Important contributions in three-dimensional, large deformation modeling of soft polymers and polymer gels with dynamic crosslinks (Polyvinyl alcohol, Polyampholyte, Ethyl acrylate copolymers, Poly(acrylamide-co-acrylic acid) hydrogel) with different microstructures and micromechanisms are limited to the quasi-static (10$^{-3}$ s$^{-1}$) to slow (10$^{-1}$ s$^{-1}$) strain rate range (\citep{Huireversiblecrosslinkmodeling, Longreversiblecrosslinkmodeling, Guoreversiblecrosslinkmodeling, FMeng(2016)Stressfreeconfigmodeling1, FMeng(2016)Stressfreeconfigmodeling2, Linreversiblecrosslinkmodeling, Venkatareversiblecrosslinkmodeling, Wangreversiblecrosslinkmodeling}). The numerical implementations of these models have also been limited to  \emph{one-dimensional, homogenous} deformation problems. Additional micromechanisms with very short relaxation timescales can contribute to the mechanical response at high strain rates. Hence, large strain material response characterization at high strain rates and corresponding constitutive models are crucial. In addition, numerical procedures to evaluate the convolution-like time integrals due to dynamic crosslink kinetics for model implementation in finite element programs are necessary to simulate complex inhomogeneous loading problems. The important gaps to be filled are (a) large strain material response characterization of PBS and (b) physics-based, three-dimensional large deformation constitutive models and their numerical implementation spanning a wide strain-rate range for PBS. 

To fill the gaps mentioned above, we have conducted the following work: (i) experimentally characterized large strain compression response of PBS up to a true strain of $\sim$125$\%$ over a wide strain rate range of 10$^{-3}$ s$^{-1}$ to 10$^{3}$ s$^{-1}$, (ii) developed a microstructure physics-motivated three-dimensional, large-deformation, rate-dependent, elastic-viscoplastic constitutive model\footnote{Although, at high strain rates (short timescales), any plastic dissipation in the polymer will lead to internal heating under adiabatic conditions, we have developed an isothermal model since the temperature rise is calculated to be very small (only a few degrees) and its effect on polymer response near ambient temperature loading can be neglected.
} for PBS and similar soft polymers and polymer gels with dynamic crosslinks. The three-dimensional model accounts for the dynamic crosslink kinetics, very high rate large strain response, and allows a continuous, smooth transition from dissipative resistance due to chain sliding at very slow strain rates to network stretching resistance at high strain rates. The model captures PBS's response over the strain rates of 10$^{-3}$ s$^{-1}$ to 10$^{3}$ s$^{-1}$. The model is validated through inhomogeneous deformation experiments, (iii) propose two types of crosslinks in PBS and identify their significantly different macroscopic relaxation timescales ($\tau$) based on the macroscopic mechanical response, namely B:O coordinate-bond dynamic crosslinks ($\tau \approx$ 3 s), and temporary entanglement lockups at high strain rates acting as crosslinks ($\tau \approx$ 0.0005 s), and (iv) outline a numerical procedure evaluating the convolution-like integrals due to crosslink kinetics for model implementation in finite element software.

\section{Experiments}
\label{Experiments}

All experiments were conducted at 23$^{\text{o}}$C. Three repetitions were performed for each experimental result reported in this paper.

\subsection{Synthesis}

PBS was synthesized by mixing hydroxy-terminated Polydimethylsiloxane entangled melt (PDMS, Sigma Aldrich 481963, 750 cSt) and
Boric Acid (BA, Sigma Aldrich B0394, ACS reagent, $\ge$ 99.5$\%$) in the ratio 160:1 by weight,
stirring the mixture and then heating it at 120$^{\text{o}}$C for 72 hrs (\citep{Seetapan2013PBSsynthesis,Tangetal(Synthesis)}). Hydroxy terminated PDMS entangled melt with a viscosity of 750 cSt has number-average and weight-average molecular mass of 30 kg/mol and 49 kg/mol, respectively, which are greater than 12 kg/mol, the critical entanglement molecular weight of PDMS.

\subsection{Rheological experiments}
\label{Rheometry}

\begin{figure}[h!]
    \begin{center}
		\includegraphics[width=\textwidth]{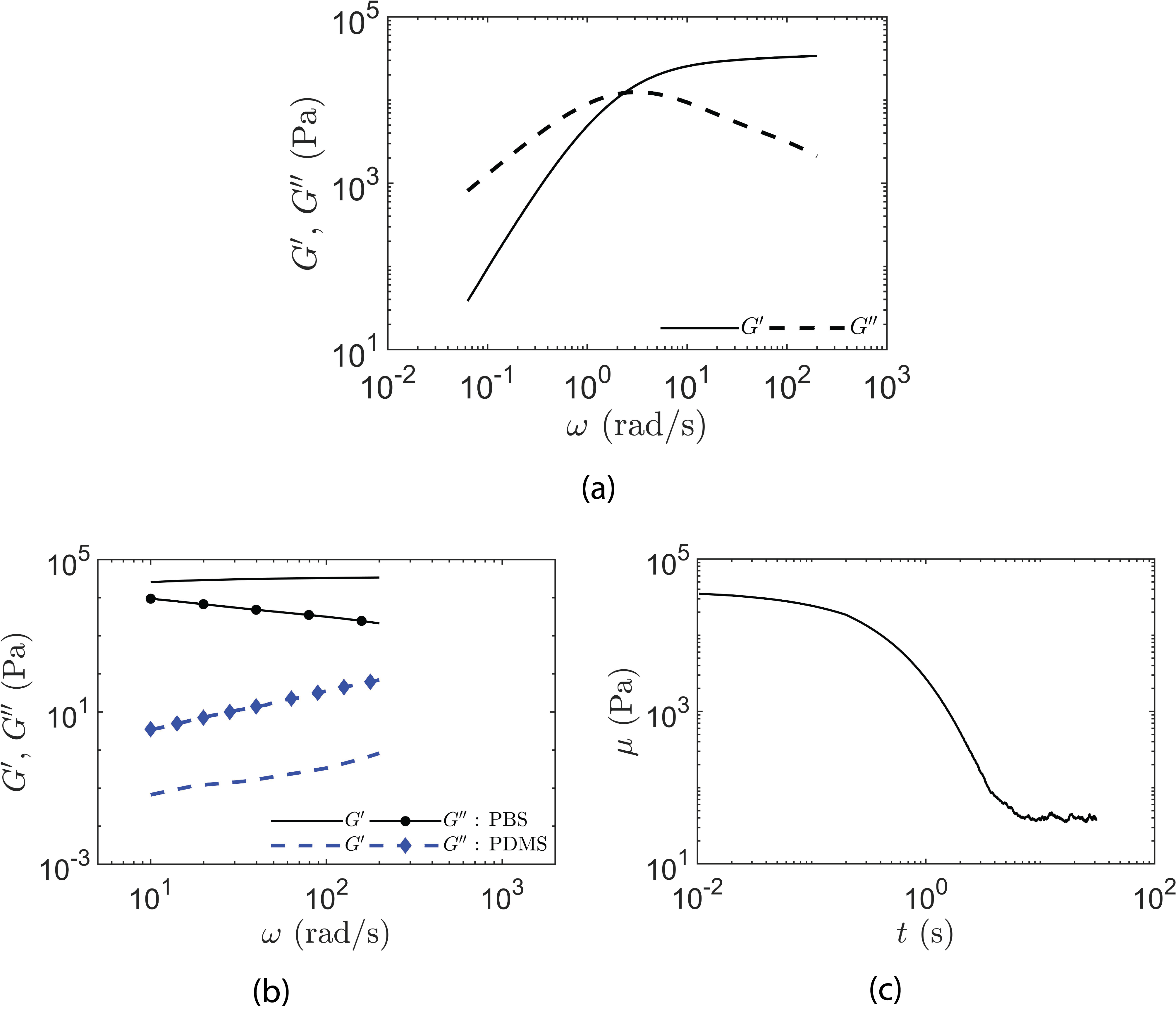}
	\end{center}
    \caption{{\small (a) Oscillatory shear response of PBS. (b) Comparison of PBS's oscillatory shear response with that of its hydroxy-terminated entangled PDMS melt precursor. The consistently
higher loss modulus than the storage modulus for PDMS entangled melt in the high frequency regime (10 - 100 rad/s) illustrates its viscous fluid-like behavior. Comparing this
with the response of PBS confirms the condensation of BA and PDMS hydroxy end groups to connect PDMS chains along their ends through Si-O-B bonds. (c) Shear stress relaxation response of PBS. The specimens were subjected to 1\% strain for all rheological experiments.
}} 
    \label{Rheometry response}
\end{figure}

Oscillatory shear and shear stress relaxation experiments were performed on an ARES G2 Rheometer to 1$\%$ strain using a cone and plate geometry (40 mm diameter and 0.04 rad cone angle). Rheological experiments are simple and effective methods to obtain basic quantitative properties of dynamically crosslinked soft polymers, e.g., macroscopic relaxation timescales of dynamic crosslinks. Results of oscillatory shear experiments on PBS are shown in Figure \ref{Rheometry response}(a). A network with only ideal, permanent crosslinks would exhibit frequency-independent storage and loss moduli (\citep{Zhao(2018)Idealreversiblenetworks}). The highly frequency-dependent storage, loss moduli, and plateauing of the storage modulus at high frequencies in Figure \ref{Rheometry response}(a) suggest the presence of dynamic crosslinks with relaxation timescales within those of the experiments. For comparison, oscillatory shear experiments were conducted on the hydroxy-terminated PDMS entangled melt precursor, the results of which are shown in Figure \ref{Rheometry response}(b). The consistently higher loss modulus than the storage modulus for PDMS entangled melt in the high frequency regime (10 - 100 rad/s) illustrates its viscous fluid-like behavior. Comparing this with the corresponding response of PBS in Figure \ref{Rheometry response}(b) confirms the condensation of BA and PDMS hydroxy end groups during the heating process to connect PDMS chains along their ends through Silicon (Si)-O-B bonds ({\citep{Tangetal(Synthesis)}}). Figure \ref{Rheometry response}(c) shows the shear stress relaxation response of PBS.

\vspace{0.05in}
\subsection{Large strain compression experiments}

\label{Compression experiments}
\vspace{0.1in}
\subsubsection{Quasi-static to moderate strain rates}
\vspace{0.05in}

Compression tests up to a true strain of 125\% were carried out using UniVert Mechanical Test System (CellScale) at true strain rates over three decades from 0.001 s$^{-1}$ to 1 s$^{-1}$. The specimens were cylindrical with a diameter of 10$\pm$2 mm and a height of 5$\pm$1 mm. The interfaces between the specimens and the platens were lubricated to prevent barreling. Figure \ref{Simple compression response}(b) and (c) show the corresponding true stress-strain response and the substantial rate-dependent stiffening. The true strain values were calculated using the standard definition: true strain = log$_e$(stretch). The specimens showed significant permanent deformation after unloading ($\approx$ 100\% at 0.001 s$^{-1}$ to $\approx$ 83\% at 1 s$^{-1}$) as shown by the unloading response curves.

\vspace{0.05in}
\subsubsection{High strain rate experiments with a SHPB setup }
\vspace{0.05in}

\begin{figure}[h!]
    \begin{center}
		\includegraphics[width=\textwidth]{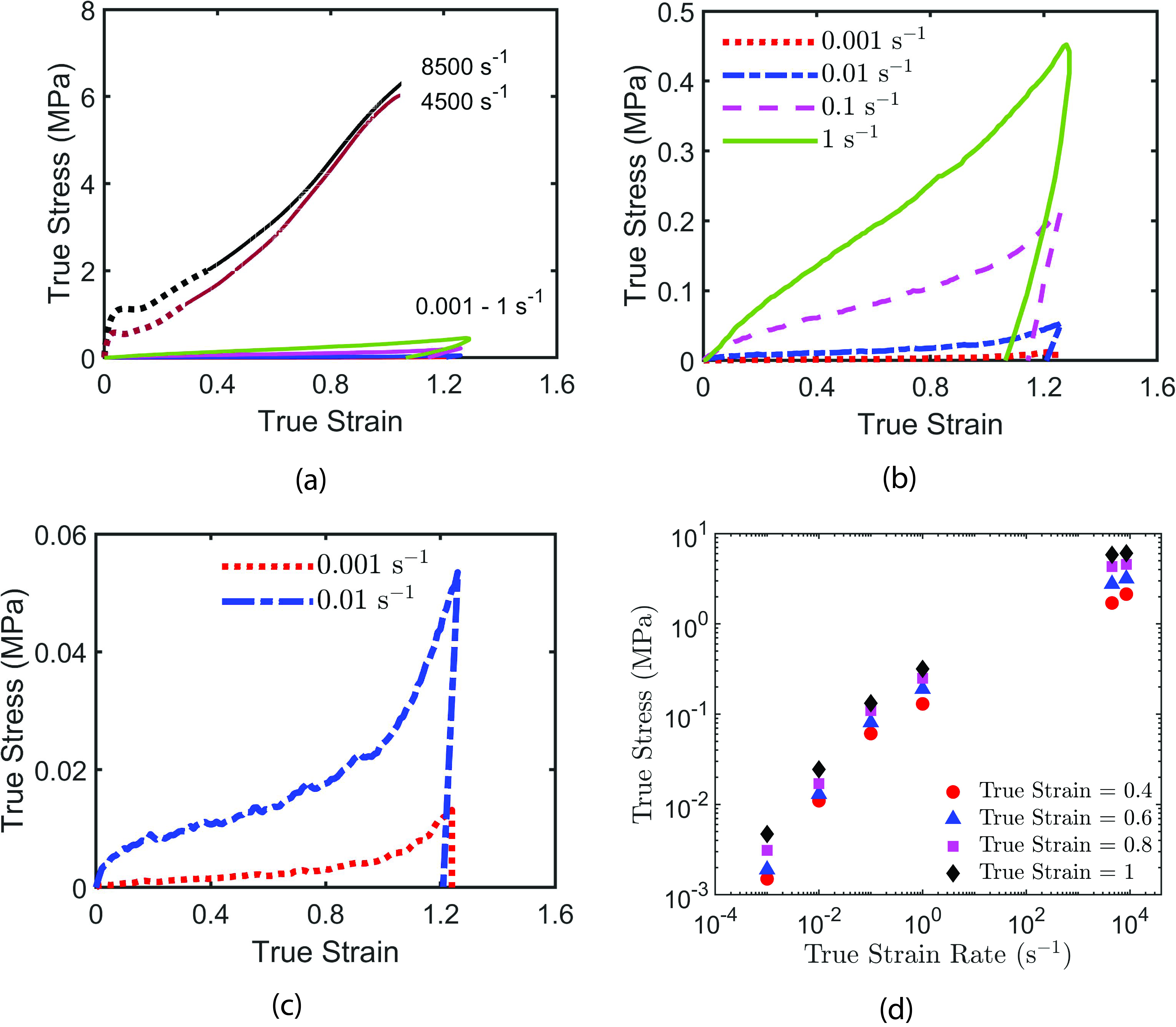}
	\end{center}
    \caption{{\small (a) Large strain compression response of PBS in the quasi-static to high (0.001 s$^{-1}$ to 8500 s$^{-1}$) strain-rate range. The dashed lines for 4500 s$^{-1}$ and 8500 s$^{-1}$ correspond to the region where the ratio between the forces on the front and back faces of the specimen was less than 0.7. (b) Zoomed-in view of the response spanning strain rates of 0.001 s$^{-1}$ to 1 s$^{-1}$. (c) Zoomed-in view of the response at the strain rates of 0.001 s$^{-1}$ and 0.01 s$^{-1}$. (d) True stress-strain rate curves at different true strain values showing a substantial rate-dependent stiffening of PBS. The stresses increase by more than three orders of magnitude as we go from the strain rate of 0.001 s$^{-1}$ to  8500 s$^{-1}$.}}
    \label{Simple compression response}
\end{figure}

SHPB setup (\citep{Kolsky}) is used to characterize the dynamic, high strain rate response of materials and structures. The specimen is sandwiched between incident and transmission bars equipped with strain gauges that measure the respective strain histories (\citep{SrivastavaSHPB1}, \citep{KHANSHPB}, \citep{SrivastavaSHPB2},  \citep{KHANSHPB2},\citep{KABIRIANSHPBIJP}, \citep{AngelaKuIJPSHPB}). A stress pulse of finite length is generated in the incident bar by impacting it with a striker bar fired from a gas gun. Once the stress pulse arrives at the specimen, part of it gets transmitted through the specimen to the transmission bar, while the remaining pulse gets reflected back into the incident bar due to impedance mismatch. Force history on both faces of the specimen and dynamic stress-strain curves can be obtained from the measured strain histories and standard analysis utilizing the theory of one-dimensional wave propagation (\citep{chen2010split}). PBS was characterized in the high strain rate regime (10$^3$ s$^{-1}$) using a SHPB setup. Due to the low stress wave impedance of PBS, a hollow aluminum transmission bar was used with inner and outer diameters of 16.51 mm and 19.05 mm, respectively. The hollow bar amplifies the transmitted stress signal compared to an equally long solid aluminum transmission bar with a diameter of 19.05 mm due to its reduced cross-sectional area. Each end of the transmission bar was plugged with aluminum end caps. Semiconductor strain gauges (Micron Instruments SS-060-033-1000PB-S1) were used on the hollow transmission bar. The solid incident bar was made of aluminum and had a diameter of 19.05 mm. Both incident and transmission bars were 1.83 m long. The specimen-incident bar and specimen-transmission bar interfaces were lubricated using a Molybdenum disulfide-based lubricant. To obtain a valid material response, it is essential to achieve dynamic force equilibrium throughout the PBS specimen\footnote{At strain rates in the order of 10$^3$ to 10$^4$ s$^{-1}$, the timescales of deformation are comparable to those of the stress wave propagation. The standard methods for evaluating material response in the form of stress-strain curves from these high strain rate experiments assume homogenous conditions throughout the specimen thickness (\citep{ShuklaSHPB,prakash2012, pinkeshpc, WEEKS2022}), which are approximately obtained usually after a significant number of stress wave reverberations through the specimen thickness. Therefore, it is essential to ensure that the specimen experiences dynamic force equilibrium (reasonably equal forces on the front and back faces of the specimen from the incident and transmission bars, respectively) over a significant portion of the deformation period for the stress-strain curves to actually represent the material response. A careful consideration of dynamic force equilibrium in characterizing soft materials at high strain rates is warranted.}. A significant number of stress wave reverberations through the specimen thickness are needed to achieve force equilibrium. Incident stress pulse duration generated using a SHPB is equal to the time taken for stress waves to make a round trip through the striker bar length (i.e., it is inversely proportional to the speed of stress waves in the striker bar) after they originate at the striker bar-incident bar interface. In low impedance materials like PBS, the relatively lower speed of stress waves implies that attaining dynamic force equilibrium using conventional high impedance striker bars (short stress pulse durations) is very difficult (\citep{ShuklaSHPB,Song2004}). A low impedance striker bar made of HYDEX\textsuperscript{\textregistered} (a type of nylon) with 0.3 m length and 12.7 mm diameter was used, which provided a sufficiently long incident stress pulse length allowing adequate time for stress to equilibrate across the specimen thickness. The cylindrical specimens had a diameter of 11.43$\pm$1 mm and a thickness of 1.04$\pm$0.1 mm. The corresponding true stress-strain response is shown in Figure \ref{Simple compression response}(a), and the true strain rate reported is an arithmetic mean over the pulse length. The difference in stress magnitudes between the high strain-rate (4500 s$^{-1}$ and 8500 s$^{-1}$) and moderate strain-rate (0.001 s$^{-1}$ - 1 s$^{-1}$) response is seen in Figure \ref{Simple compression response}(a)\footnote{Experimental methods to accurately characterize the stress-strain response of soft materials like PBS in the intermediate strain rate range of 10$^{1}$ s$^{-1}$ to 10$^{2}$ s$^{-1}$ continue to be a challenge and is an active area of research.}. To get a quantitative sense, true stress-strain rate curves at different strain levels are shown in Figure \ref{Simple compression response}(d). The stresses increase by slightly more than three orders of magnitude as we go from the strain rate of 0.001 s$^{-1}$ to 8500 s$^{-1}$.

\section{Theory and constitutive model}
\label{Specialization}

The key features needed to be captured by a three-dimensional, large deformation constitutive model for PBS are (i) nonlinear large strain response, (ii) kinetics of the dynamic crosslinks involved, and (iii) significant rate-dependent stiffening of the material response. 

We base our theory on the classical Kroner ({\citep{Kroner(1959)}})-Lee (\citep{Lee(1969)}) multiplicative decomposition of the deformation gradient $\textbf{F}$ into elastic and plastic parts $\textbf{F}^e$ and $\textbf{F}^p$ which is commonly used in elastic-viscoplastic and visco-elastic constitutive theories (\citep{BOYCE1988}, \citep{Ekuhl1}, \citep{Ekuhl2}, \citep{Mjmetastablesteels}, \citep{FEFPIJP}, \citep{MahrezIJPFeFp}, \citep{SijunplasticityPINN}, \citep{LINIJPFeFp}, \citep{WANGIJPFeFp})

\begin{equation}
	\mathbf{F}=\mathbf{F}^e \mathbf{F}^p \:\:\:\:  \text{with} \:\:\:\:  \text{det}\textbf{F}^e > 0 \:\:\:\:  \text{and} \:\:\:\:  \text{det}\textbf{F}^p > 0.
	\label{Multiplicative decomposition of F}
\end{equation}

We utilize the parallel ``multi-mechanism" generalization of the multiplicative decomposition in \eqref{Multiplicative decomposition of F} that accounts for various rate-dependent microscopic relaxation mechanisms (\citep{BergstromBoyce(1998)}, \citep{ANAND2009}, \citep{AMES2009}, \citep{Srivastava2011Thermo}, \citep{Srivastavashapememory}, \citep{LIMultmechIJP}, \citep{GONZALEZMultimechIJP}, \citep{MAHJOUBIMultmechIJP}, \citep{HAOMultmechIJP}, \citep{XIAOMultmechIJP}), i.e.

\begin{equation}
    \mathbf{F}=\mathbf{F}^{e(\alpha)} \mathbf{F}^{p(\alpha)} \:\:\:\:  \text{with} \:\:\:\:  \text{det}\textbf{F}^{e(\alpha)} > 0 \:\:\:\:  \text{and} \:\:\:\:  \text{det}\textbf{F}^{p(\alpha)} > 0, \:\:\:\: \alpha = 1,....,M,
    \label{Multimechanism generalization of Fe Fp decomposition}
\end{equation}

where each $\alpha$ represents one of the $M$ mechanisms acting in parallel. $M$ denotes the total number of different physical mechanisms contributing to the polymer's mechanical response. 
 For a given material point $\mathbf{X}$, with invertible $\textbf{F}^{e(\alpha)}$ and $\textbf{F}^{p(\alpha)}$,  

\begin{equation}
    \mathscr{M}_{\mathbf{X}}^{(\alpha)}\overset{\text{def}}{=} \text{range of} \: \textbf{F}^{p(\alpha)}(\textbf{X}) = \text{domain of} \: \textbf{F}^{e(\alpha)}(\textbf{X}) 
    \label{Intermediate structural space}
\end{equation}

is referred to as the intermediate structural space at a material point $\mathbf{X}$ for a given $\alpha$ which is a purely mathematical construct. We assume PBS as an isotropic material. An isothermal constitutive model is developed for applications involving near room temperature conditions owing to a very small estimated temperature rise (only a few degrees Celsius) due to inelastic dissipation even for the strain rate of 8500 s$^{-1}$. The physically motivated discussions in the following Sections \ref{B:O dative dynamic crosslinks} and \ref{Entanglements acting as dynamic crosslinks} help identify the dominant mechanisms of deformation in PBS.   

\vspace{0.05in}
\subsection{Dynamic, reversible crosslinks}
\label{Crosslink kinetics}

\vspace{0.05in}
\subsubsection{B:O coordinate-bond dynamic crosslinks, rheological test predictions}
\vspace{0.05in}

\label{B:O dative dynamic crosslinks}

During the synthesis of PBS, hydroxy end groups on Boric Acid (BA) and hydroxy-terminated Polydimethylsiloxane (PDMS) chains in the entangled melt condense to form Silicon(Si)-O-B bonds connecting the ends of hydroxy-terminated PDMS chains through boron atoms and form longer chain molecules ({\citep{Tangetal(Synthesis)}}). The electron-deficient p orbitals of these connecting boron atoms can further receive a pair of electrons from oxygen atoms in the dimethylsiloxane backbone of neighboring end-connected PDMS chains to form dynamic coordinate B:O bonds that act as dynamic crosslinks with the ability to relax and reconnect (\citep{Lietal(2014)B:Ocrosslinkref}). We use ``subchains" to denote parts of chain molecules between two elastically active crosslinks (connected to the elastically active network). We use the superscript ``$bd$” to indicate quantities associated with subchains between B:O coordinate-bond dynamic crosslinks. We denote the referential number density of chain segments currently in their ``subchain" state (S) corresponding to dynamic B:O coordinate-bond crosslinks as $n^{bd}_s$ and that of chain segments which can be in the (S) state but currently are in the ``not subchain" state (NS) as $n^{bd}_{ns}$. The underlying dynamic, reversible reactions can be represented as

\begin{equation}
   n^{bd}_{ns} \:\:\:\: (\mathrm{NS}) \underset{k^{bd}_{ns}}{\stackrel{k^{bd}_{s}}{\rightleftharpoons}} (\mathrm{S}) \:\:\: n^{bd}_{s},
   \label{B:O reversible crosslink reaction}
\end{equation}

where $k^{bd}_{s}$ and $k^{bd}_{ns}$ are the forward and backward reaction rate parameters. Assuming first-order kinetics to model the reactions, we have

\begin{equation}
    \dot{n}^{bd}_s = k^{bd}_{s} n^{bd}_{ns} - k^{bd}_{ns} n^{bd}_s. 
    \label{B:O reversible crosslink first order kinetics} 
\end{equation}

We now focus on the shear stress relaxation data in Section \ref{Rheometry}. We assume that (i) the new crosslinks and subchains are formed in their stress-free, relaxed configurations,  (ii) B:O coordinate-bond crosslinks are the dominant type of dynamic crosslinks with relaxation timescales within those of the shear stress relaxation and oscillatory shear experiments, (iii) relaxation of B:O coordinate-bond dynamic crosslinks is the dominant relaxation mechanism in the timescales of the rheology experiments, (iv) the reaction rate parameters are approximately constant under the applied small strain loadings, and (v) we start from a spatially homogenous, stress-free, equilibrium state of the kinetics at $t$=0, i.e., $n^{bd}_s(0)$ and $n^{bd}_{ns}(0)$ related through $n^{bd}_{ns}(0) = \bigl[k^{bd}_{ns}/k^{bd}_s\bigr] n^{bd}_s(0)$. 

For the case of stress relaxation, the subchains formed after the strain ramp do not stretch and do not carry any load during the hold period. Further, following the theory of rubber elasticity (\citep{Treolar}) and assumption (iii), the shear stress relaxation modulus $\mu(t)$ is given by 

\begin{equation}
    \mu(t) = n^{bd}_{load}(t)\text{k}_{\text{B}}\theta,
    \label{Affine network theory ground state shear modulus}
\end{equation}

where k$_{\text{B}}$ is the Boltzmann constant and $n^{bd}_{load}$ denotes the referential number density of load bearing subchains during the hold period\footnote{Since the strain ramp for the stress relaxation experiment is faster than $k^{bd}_{ns}$ and $k^{bd}_{s}$, all $n^{bd}_s(0)$ subchains at $t$=0 will be stretched to 1\% strain at the end of the strain ramp. Hence, $n^{bd}_{load}(t_{ramp}) = n^{bd}_{s}(0)$ and all $n^{bd}_{load}(t)$ load bearing subchains at any time instant $t\ge t_{ramp}$ are at 1\% strain.}. Considering only the stress decay (relaxation) corresponding to the conversion of load bearing subchains to the not-subchain state, the form of {\eqref{B:O reversible crosslink first order kinetics}} for the evolution of $n^{bd}_{load}$ during the hold period ($t\ge t_{ramp}$) reduces to 

\begin{equation}
    \dot{n}^{bd}_{load} = -k^{bd}_{ns}n^{bd}_{load} \:, \: n^{bd}_{load}(t) = n^{bd}_s(0)e^{-k^{bd}_{ns}t} \:\: (\textrm{for } \; t\ge t_{ramp}). 
    \label{Decaying first order kinetics (relaxed newborn crosslinks)}
\end{equation}
Substituting \eqref{Decaying first order kinetics (relaxed newborn crosslinks)} in \eqref{Affine network theory ground state shear modulus}, we get

\begin{equation}
    \mu(t) = \mu(0) e^{-k^{bd}_{ns}t},
    \label{Final exponentially decaying ground state shear modulus}
\end{equation}

and

\begin{equation}
    \mu(0) = n^{bd}_{s}(0) \text{k}_{\text{B}} \theta.
    \label{Ground state shear modulus at t=0}
\end{equation}

\begin{figure}[h!]
    \begin{center}
		\includegraphics[width=\textwidth]{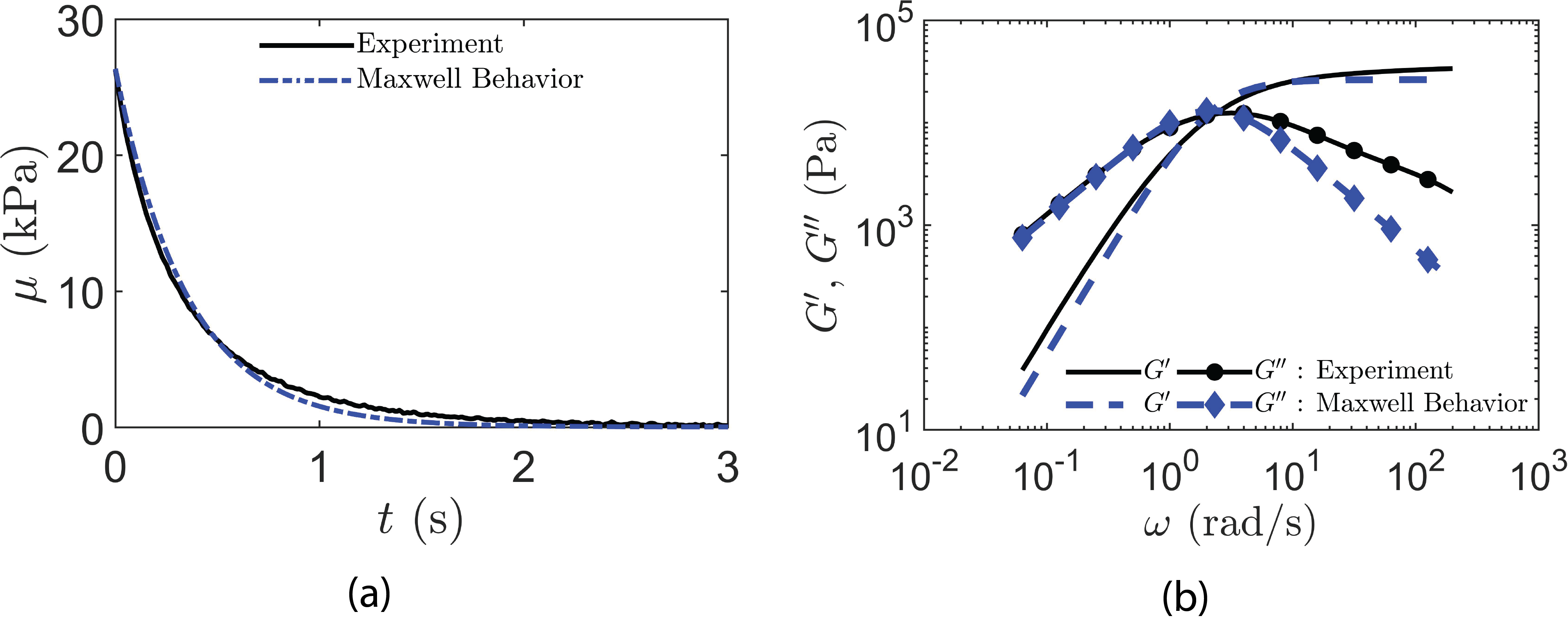}
	\end{center}
    \caption{{\small (a) Fit of the linear Maxwell-like shear stress relaxation response of PBS at small strains assuming first-order kinetics for B:O coordinate-bond dynamic crosslinks. The new subchains are formed in their stress-free, relaxed configurations. (b) Oscillatory shear loading predictions of $G^{'}$ and $G{''}$ considering the Maxwell-like behavior with shear stress relaxation test fitted parameters $n^{bd}_s(0)$, $k^{bd}_{ns}$.}} 
    \label{Maxwell model predictions}
\end{figure}

As seen in Figure \ref{Maxwell model predictions}(a), equations \eqref{Final exponentially decaying ground state shear modulus}, \eqref{Ground state shear modulus at t=0} fit the shear stress relaxation experimental data well with $n^{bd}_s(0) = 6.4 \: \text{x} \: 10^{24} \: \text{m}^{-3} \: \text{and} \: k^{bd}_{ns} = 0.35 \: \text{s}^{-1}$. Further, \eqref{Final exponentially decaying ground state shear modulus} is identical to the expression for shear stress relaxation modulus of a Maxwell element,

\begin{equation}
    \mu(t) = \mu(0) \: e^{-\dfrac{t}{\tau}}, 
    \label{Maxwell model shear modulus exponential decay}
\end{equation}

and 

\begin{equation}
    \tau = \dfrac{1}{k^{bd}_{ns}},
    \label{Equivalence of timescale parameters, Maxwell model}
\end{equation}

quantifying the B:O coordinate-bond dynamic crosslink \emph{macroscopic} relaxation timescale. This implies that PBS will behave like a Maxwell element in other cases of small strain loadings as well. Considering the linear viscoelastic Maxwell element-like behavior, the storage and loss moduli $G^{'}(\omega)$, $G^{''}(\omega)$ for a sinusoidal loading corresponding to a frequency of $\omega$ for PBS are given as

\begin{equation}
    G^{'}(\omega) = \mu(0)\dfrac{\omega^{2}}{\omega^{2} + (2\pi k^{bd}_{ns})^{2}}, \:\: G^{''}(\omega) = \mu(0)\dfrac{2 \pi \omega k^{bd}_{ns}}{\omega^{2} + (2\pi k^{bd}_{ns})^{2}}.
    \label{Storage and loss moduli (PBS)}
\end{equation}
 
Response predictions for the oscillatory shear loading obtained using \eqref{Storage and loss moduli (PBS)} and $\mu$(0), $k^{bd}_{ns}$ from fit to the shear stress relaxation response are in good agreement with the experimental results as seen in Figure \ref{Maxwell model predictions}(b). Fit of the shear stress relaxation data and reasonably good prediction for the oscillatory shear loading support the assumption of B:O coordinate-bond dynamic crosslink relaxation being the dominant relaxation mechanism in the shear stress relaxation and oscillatory shear experimental timescales. This also supports the assumptions of new crosslinks and subchains being formed in their stress-free relaxed configurations and the reaction rate parameters being approximately constant under the applied small strain loading. When the loading timescales are such that $t_{\text{load}} \ll 1/k^{bd}_{ns}$, the dynamic B:O coordinate-bond crosslinks do not have sufficient time to dissociate. Consequently, PBS will resemble a network with ``permanent" crosslinks. Therefore, PBS will behave like a Maxwell element up to frequencies near the crossover frequency $\omega_{c}$ = $2\pi k^{bd}_{ns}$ and will transition to ``permanent" network-like behavior with B:O coordinate-bond dynamic crosslinks for frequencies beyond $\omega_c$. The plateauing of $G^{’}$ after the crossover frequency, when a permanent like network is approached, suggests approximately constant number density of load bearing subchains, i.e., negligible formation of new subchains which are stretched in these relatively short timescales. This indicates that $k^{bd}_s$ is comparable to or smaller than $k^{bd}_{ns} = 0.35 \: \text{s}^{-1}$.

\vspace{0.05in}
\subsubsection{Temporary entanglement lockups at high strain rates}
\label{Entanglements acting as dynamic crosslinks}
\vspace{0.05in}

Using the fitted parameter $n^{bd}_s(0)$ and {\eqref{Ground state shear modulus at t=0}}, we get a ground state shear modulus $\mu(0)$ of approximately 26 kPa. For loading timescales $t_{load} \lll 1/{k^{bd}_{ns}}$, the subchain kinetics can be neglected, i.e., a ``permanent" - like network is approached. B:O coordinate-bond dynamic crosslinks with $\mu(0)$ = 26 kPa is not sufficient to explain the relatively substantially larger stress levels observed at the high strain rate of 8500 s$^{-1}$ ($t_{load}$ = 1/(8500 s$^{-1}$) $\lll 1/k^{bd}_{ns}$) in Figure \ref{Simple compression response}(a). This suggests a possibility of the presence of an additional type of dynamically crosslinked network with dynamic crosslink relaxation timescales much shorter than those involved in the oscillatory shear experiments. 

Previous work, including molecular dynamics simulations ({\citep{MDentangledmelts}}) and macroscopic experiments ({\citep{Macroscopicexpentangledmelts}}) have suggested that temporary entanglement lockups can occur in entangled polymer melts at high strain rates through tightening of interchain twists and knots which are randomly created and destroyed. At low strain rates, the twists and knots can disentangle, and chains can slip past each other and flow similar to the tube model predictions. The twists and knots tightening up at high strain rates enable the transition of entangled polymer melts from viscous flow to rubbery-like behavior at high strain rates. The entanglement lockups which temporarily occur at high strain rates can be thought of as reversible crosslinks with very short relaxation timescales. PBS shows behavior similar to entangled polymer melts at slow strain rates. Hence, we suggest an additional network stretching resistance coming from temporary entanglement lockups at high strain rates acting as crosslinks with very short relaxation timescales\footnote{PBS without entanglements (synthesized using PDMS melts with molecular weights below the entanglement weight of PDMS) show oscillatory shear response similar to the results in Figure \ref{Rheometry response}(a) (\citep{Tangetal(Synthesis),Kurkinetal(Synthesis)}) . This indicates that the relaxation timescale of 0.35 s$^{-1}$ is related to the macroscopic relaxation of B:O coordinate-bond dynamic crosslinks.}. 

\begin{figure}[h!]
	\begin{center}
		\includegraphics[width=\textwidth]{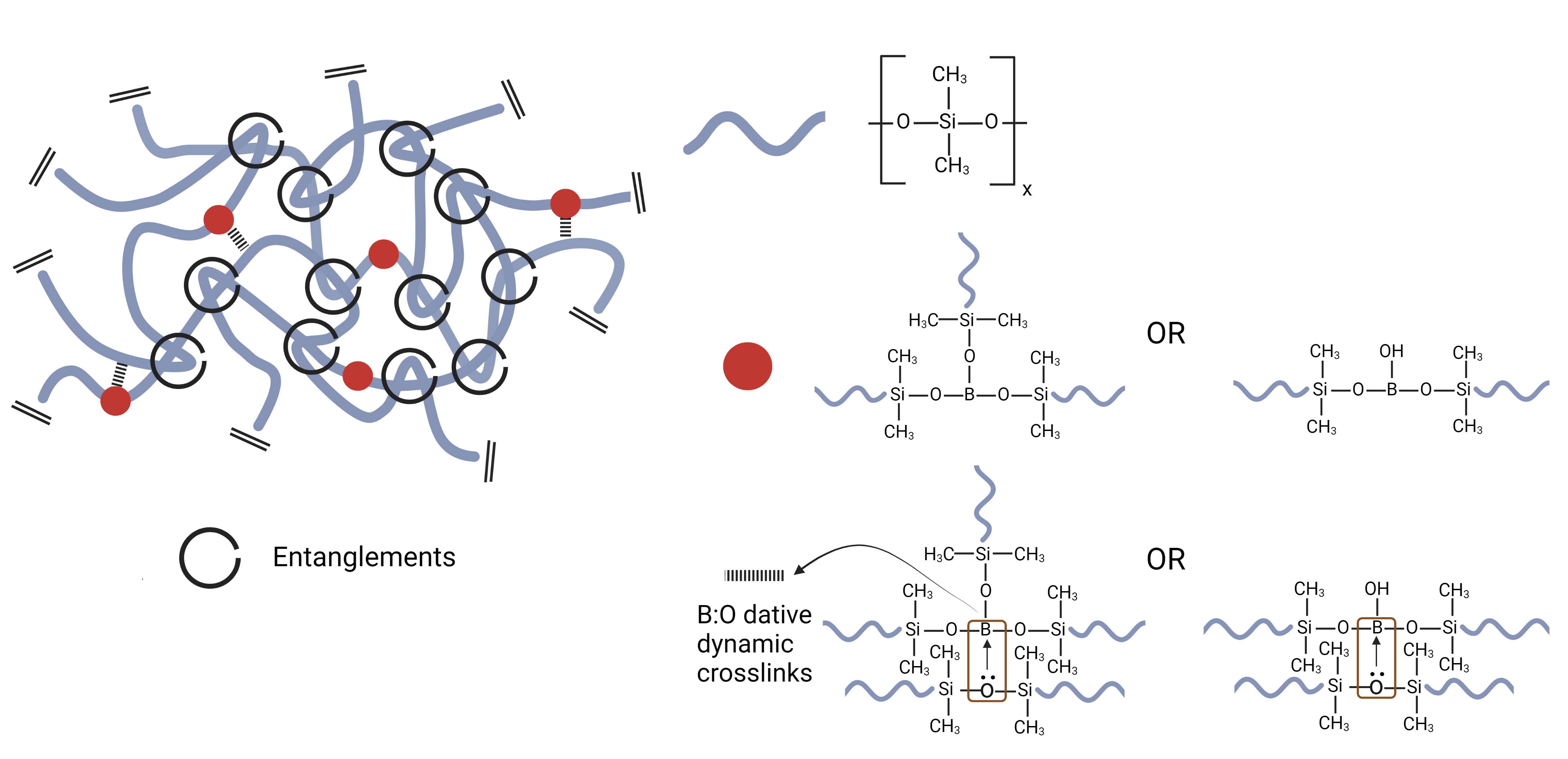}
	\end{center}
	\caption{{\small 
		Schematic of B:O coordinate-bond dynamic crosslinks and entanglements in PBS.}}
	\label{reversible crosslink schematic}
\end{figure}

We use the superscript ``$el$" to denote quantities associated with subchains between entanglement lockups. Figure {\ref{reversible crosslink schematic}} shows a schematic representation of the B:O coordinate-bond dynamic crosslinks and entanglements.  

\vspace{0.05in}
\subsubsection{Force dependence of reaction rate parameters}
\label{Force dependence of reaction rate parameters}
\vspace{0.05in}

\begin{figure}[h]
	\begin{center}
		\includegraphics[width=\textwidth]{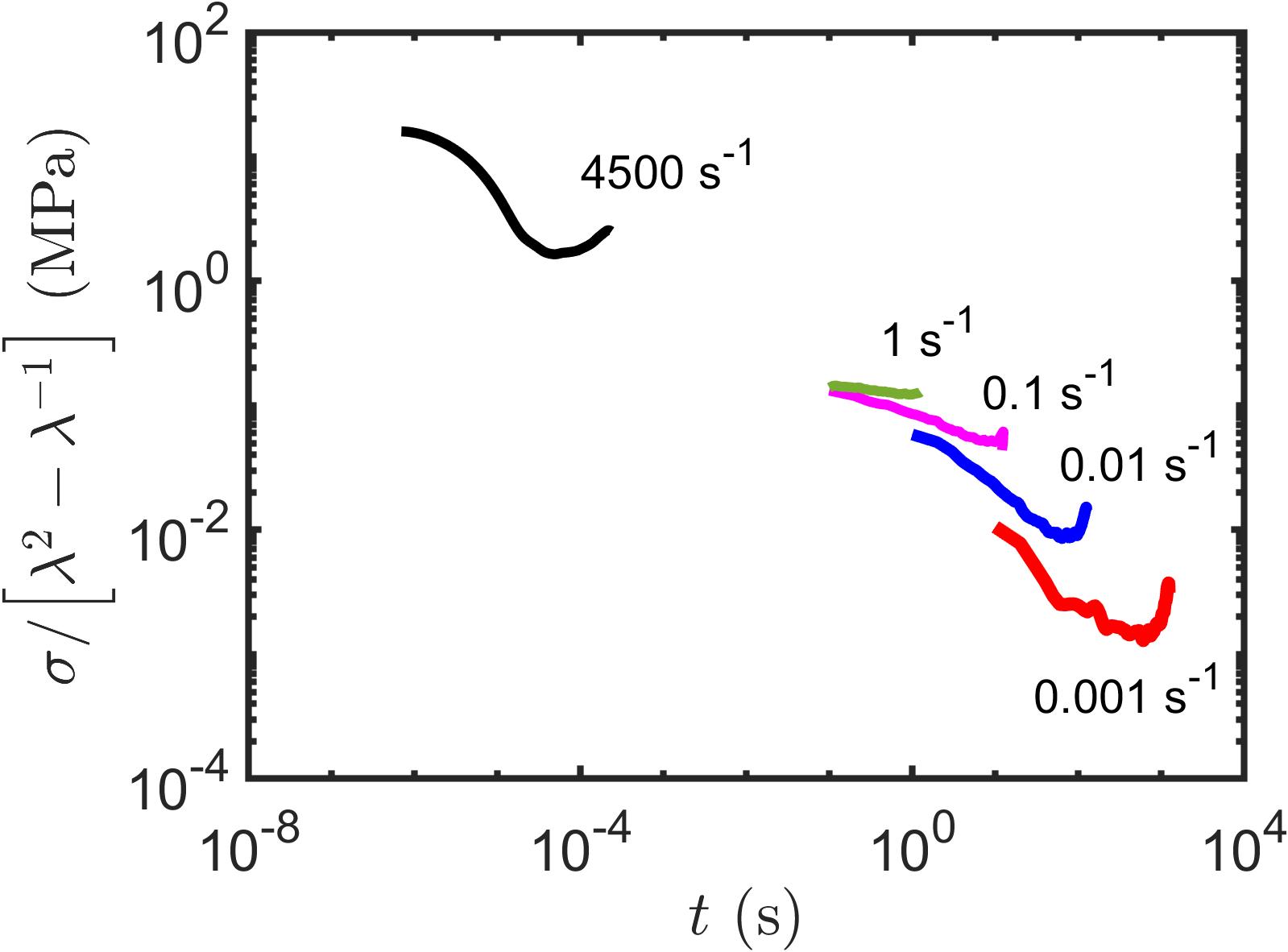}
	\end{center}
	\caption{{\small 
			(a) Reduced stress, $\sigma_r = \dfrac{\sigma}{\lambda^2 - \lambda^{-1}}$ vs compressive stretch, $\lambda$ curves for strain rates from 0.001 s$^{-1}$ to 8500 s$^{-1}$ obtained using the compression experimental data in Section \ref{Compression experiments}. $\sigma_{r}$ is approximately constant with $\lambda$ indicating weak dependence of the reaction rate parameters on the force for the experimental stretch range in this study. (b) $\sigma_{r}-t$ curves for the wide range of strain rates. The curves are close to overlapping each other for the strain rates of 0.001 s$^{-1}$ to 1 s$^{-1}$ till moderate stretch levels reflecting the approximate force independence of the reaction rate parameters. 
	}}
	\label{Reduced stress vs time}
\end{figure}

The reaction rate parameters $k_{ns}$ and $k_{s}$ experienced by subchains formed at $t^{'}$ ($0<t^{'}$) and surviving till the current time instant $t$ will, in general, depend on the forces experienced by them ({\citep{Bellkinetics}}). The effect of forces acting on the subchains on the reaction rate parameters can be qualitatively understood using a simple analysis similar to the work of Mayumi and co-workers ({\citep{Mayumi2013}}). The subchains formed at $t$=0 and surviving till $t$ would experience the largest stretch levels equal to the total imposed stretch and hence the largest forces. With the Neo-Hookean model as an example, a reduced stress measure under uniaxial loading for the subchains formed at $t=0$ and surviving till $t$ is defined as $\sigma_r$ = $\sigma/(\lambda^2-\lambda^{-1})$. Here $\sigma$ is the true stress, $\lambda$ is the total imposed stretch and $\sigma_r$ is a function of the reaction rate parameters. Figure \ref{Reduced stress vs time}(a) and (b) are $\sigma_{r}-\lambda$ and $\sigma_{r}-t$ curves obtained using the compression experimental data in Section \ref{Compression experiments}. Figure \ref{Reduced stress vs time}(a) shows that $\sigma_{r}$ is approximately constant with $\lambda$, indicating a weak dependence of the reaction rate parameters on the force for the experimental stretch range in this paper. This translates to $\sigma_{r}-t$ curves for strain rates from 0.001 s$^{-1}$ to 1 s$^{-1}$ in Figure \ref{Reduced stress vs time}(b) being close to overlapping each other till moderate stretch levels before finite extensibility of the chains and other strain-stiffening non-linearities not accounted for by the Neo-Hookean model come into the picture. Consequently, we make a reasonable assumption of the reaction rate parameters being approximately constant in the stretch range of our experiments. 
 
\vspace{0.1in}

\subsection{Mechanisms of deformation}
\label{Mechanisms of deformation}

We propose \emph{three parallel mechanisms} of deformation $A$, $B1$ and $B2$ to describe the experimentally observed response of PBS over the strain rate range of 10$^{-3}$ s$^{-1}$ to 10$^{3}$ s$^{-1}$.

(a) \textbf{Intermolecular Resistance} ($A$): Elastic and dissipative resistance due to intermolecular interactions between neighboring molecules is represented by mechanism $A$. It can be considered as a nonlinear spring in series with a dashpot.

The network resistances are modeled by mechanism $B$, which is a combination of two parallel mechanisms $B1$ and $B2$.

(b) \textbf{Network Resistance} ($B$): Resistance due to changes in the free energy upon stretching of the elastically active networks, i.e., stretching of  constituent subchains between crosslinks corresponding to 

--- \textbf{B:O coordinate-bond dynamic crosslinks} represented by mechanism ($B1$), and 

--- \textbf{Temporary entanglement lockups at high strain rates acting as crosslinks} represented by mechanism ($B2$).\\

Network mechanisms $B1$ and $B2$ are parallel to mechanism $A$. At very low strain rate mechanical deformations, the networks due to dynamic crosslinks between long chains have a negligible effect on the response, and PBS macroscopically behaves as an entangled polymer melt where the dissipative molecular resistance (mechanism $A$) due to chains sliding past each other provides the primary contribution to the mechanical response. As the strain rate increases and networks due to dynamic crosslinks begin to form, the \emph{rate-dependent} network resistance (networks $B1$ and $B2$) starts to contribute dominantly to the mechanical response and the contribution of mechanism $A$ to the overall rate-dependent stress response becomes less significant. The behavior of dynamically crosslinked soft polymers described above is substantially different than that of rubbers with permanent
crosslinks. 
 
The total referential free energy density can be additively decomposed into contributions from each mechanism of deformation as

\begin{equation}
    \psi_R \:\: = \underbrace{{\psi}^{(A)}}_\text{Intermolecular resistance} + \underbrace{{\psi}^{(B1)} + {\psi}^{(B2)}}_\text{Network resistance}.
    \label{Multimechanism generalization of free energy}
\end{equation}

The total Cauchy stress \textbf{T} following the parallel multi-mechanism generalization and additive decomposition of the total free energy density in \eqref{Multimechanism generalization of free energy} is the sum of the contributions from each of the parallel mechanisms i.e. 

\begin{equation}
    \mathbf{T} = \mathbf{T}^{(A)} + \mathbf{T}^{(B1)} + \mathbf{T}^{(B2)},
    \label{Multimechanism generalization of Cauchy stress}
\end{equation}

where $\mathbf{T}^{(A)}, \mathbf{T}^{(B1)}$ and $\mathbf{T}^{(B2)}$ are the Cauchy stress contributions from mechanisms $A, B1$ and $B2$, respectively.

\vspace{0.1in}
\subsection{Constitutive equations for intermolecular mechanism $A$}
\label{modeling A}
\vspace{0.1in}
\subsubsection{Free energy density, Cauchy stress and Mandel stress:}
\vspace{0.05in}
With $\mathbf{E}^{e(A)}=\mathrm{ln} \mathbf{U}^{e(A)}$ denoting an elastic logarithmic strain measure, where $\mathbf{U}^{e(A)}$ is the right elastic stretch tensor, we adopt for $\psi^{(A)}$ the following special form for moderately large elastic stretches (\citep{Anand(1979)Moderatestretchfreeenergy1,Anand(1986)Moderatestretchfreeenergy2})
\begin{equation}
    \psi^{(A)} = {\bar{\psi}}^{(A)} (\mathbf{E}^{e(A)}) = G \left|\mathbf{E}^{e(A)}_0\right|^2 +  \frac{1}{2} K \left( \mathrm{tr} \mathbf{E}^{e(A)}\right)^2, \footnote{\textbf{A}$_0$  \text{denotes the deviatoric part of a tensor} \textbf{A}.}
    \label{Final free energy (Mechanism A)}
\end{equation}

where $G>0$, $K>0$ are the shear and bulk modulus respectively. The contribution $\mathbf{T}^{(A)}$ corresponding to the special free energy function considered in \eqref{Final free energy (Mechanism A)} is given by 
\begin{equation}
	 \mathbf{T}^{(A)}=J^{-1} \mathbf{R}^{e(A)} \mathbf{M}^{e(A)} \mathbf{R}^{e(A) \top}, \: \mathbf{M}^{e(A)}=\frac{\partial {\bar{\psi}}^{(A)} (\mathbf{E}^{e(A)})}{\partial \mathbf{E}^{e(A)}}=2 G \mathbf{E}^{e(A)}_0 + K \left(\mathrm{tr} \mathbf{E}^{e(A)} \right) \mathbf{1}. 
	\label{Cauchy stress and Mandel stress relation (Mechanism A)}
\end{equation} 

where $\mathbf{M}^{e(A)}$ is the Mandel stress driving the plastic flow and is defined with respect to the local intermediate structural space for mechanism $A$.

\subsubsection{Flow rule}
\vspace{0.05in}

The evolution equation and initial condition for $\textbf{F}^{p(A)}$ with the assumption of isotropy and resulting plastic irrotationality, as well as the standard assumption of the body initially being in its virgin state, are given as
\begin{equation}
	\dot{\mathbf{F}}^{p(A)}=\mathbf{D}^{p(A)} \mathbf{F}^{p(A)}
    \label{Evolution equation of Fp(A)}, \: \: \mathbf{F}(\mathbf{X},0) = \mathbf{F}^{e(A)}(\mathbf{X},0) = \mathbf{F}^{p(A)}(\mathbf{X},0) = \mathbf{1}.
\end{equation}

 Assuming codirectionality of plastic flow, the plastic stretching tensor $\textbf{D}^{p(A)}$ can be expressed as

\begin{equation}
	\mathbf{D}^{p(A)}= \nu^{p(A)} \left(\frac{\mathbf{M}^{e(A)}_0}{2 \overline{\tau}^{(A)}} \right), \hspace{0.1in} \nu^{p(A)}\overset{\text{def}}{=}\sqrt{2} \left|\mathbf{D}^{p(A)} \right|, \hspace{0.1in} \overline{\tau}^{(A)}\overset{\text{def}}{=}\dfrac{1}{\sqrt{2}} \left|\mathbf{M}^{e(A)}_0 \right|
    \label{Plastic stretching tensor, equivalent plastic shear strain rate (Mechanism A)}
\end{equation}

where $\nu^{p(A)}$ is an equivalent plastic shear strain rate and $\overline{\tau}^{(A)}$ is an equivalent shear stress. We also define $\nu \overset{\text{def}}{=} \sqrt{2}\left|\textbf{D}_0\right|$ as the total equivalent shear strain rate and $\overline{p}^{(A)} \overset{\text{def}}{=} -\dfrac{1}{3} \mathrm{tr}\mathbf{M}^{e(A)}$ as the mean normal pressure. The distortional parts of $\textbf{F}$, $\textbf{F}_{dis}$, the right Cauchy Green deformation tensor ($\textbf{C}$), $\textbf{C}_{dis}$ and the effective total distortional stretch, $\overline{\lambda}$ are defined as  

\begin{equation}
	\mathbf{F}_{dis}=J^{-1/3} \mathbf{F}, \hspace{0.1in} \mathrm{det} \mathbf{F}_{dis}=1 \:\: ; \:\: \mathbf{C}_{dis}= (\mathbf{F}_{dis})^\top \mathbf{F}_{dis}=J^{-2/3} \mathbf{C} \:\: ; \:\: \overline{\lambda} \overset{\text{def}}{=} \sqrt{\dfrac{\mathrm{tr} \mathbf{C}_{dis}}{3}}
    \label{Distortional part of F and C definition}
\end{equation}

We choose the flow function for the evaluation of $\nu^{p(A)}$ in the following (similar to \citep{Srivastava2011Thermo}) form 

\begin{equation}
	 \nu^{p(A)}=\begin{cases}
\begin{aligned}
			&0 \hspace{0.2in} \hspace{0.1in} &&\text{if}  \hspace{0.1in}\bar{\tau}^{(A)}_e \leq 0,\\
			&\nu_0 \left[\mathrm{sinh} \left({\frac{\overline{\tau}^{(A)}_e}{S}} \right) \right] ^{1/{m}}  && \text{if} \hspace{0.1in} \bar{\tau}^{(A)}_e>0,
\end{aligned}
	\end{cases}
\label{Final overstress equivalent plastic shear strain rate relation}
\end{equation}

where
\begin{equation}
	\overline{\tau}^{(A)}_e \overset{def}{=} \overline{\tau}^{(A)} -\alpha_{p} \overline{p}^{(A)},
    \label{Effective equivalent shear stress driving the plastic flow definition}
\end{equation}

denotes a net equivalent shear stress for plastic flow. Here $\alpha_{p} > 0$ is a pressure sensitivity parameter, $\nu_0$ is a reference strain rate and $m$ is a strain rate sensitivity parameter. The internal variable $S$ models the dissipative friction-like resistance due to the sliding chains. The evolution of $S$ is given by the following differential equation

\begin{equation}
	\dot{S}=h \left(\overline{\lambda} -1 \right) \nu, \hspace{0.2in} \text{with initial value} \hspace{0.1in} S(\mathbf{X},0)=S_0 > 0,
	\label{Evolution equation for internal variable}
\end{equation}

where $h$ is a stress-dimensioned material parameter.

\vspace{0.1in}
\subsection{Constitutive relations for network stretching mechanisms $B1$ and $B2$}
\label{modeling B1, B2}
\vspace{0.05in}
\subsubsection{Free energy and Cauchy stress}
\label{Free energy B}
\vspace{0.05in}

\emph{The rate-dependent network stretching mechanisms B1 and B2 contribute dominantly to the overall polymer response as the strain rates start to increase.} Since we have already accounted for volumetric elastic energy in $\psi^{(A)}$, we do not allow for volumetric energy in $\psi^{(B1)}$ and $\psi^{(B2)}$. We consider free energy functions of the special form

\begin{equation}
	{\psi}^{(B1)}={\hat{\psi}}^{(B1)} \left( \mathbf{C}_{dis} \right), \:\: {\psi}^{(B2)}={\widetilde{\psi}}^{(B2)} \left( \mathbf{C}_{dis} \right).
    \label{Free energy distortional representation (Mechanism B)}
\end{equation}

With the effective total distortional stretch, $\overline{\lambda} \overset{\text{def}}{=} \sqrt{\dfrac{\mathrm{tr} \mathbf{C}_{dis}}{3}}$, we utilize (with modification) the eight-chain free energy density function denoted by $\psi$ (\citep{ArrudaBoyce(1993), BergstromBoyce(1998), ArrudaBoycereview, JerryQiBoyce}) for this purpose 
\begin{equation}
	\psi= \mu \lambda^{2}_L \left[ \left(\dfrac{\overline{\lambda}}{\overline{\lambda}_L}\right)\beta + ln\left(\dfrac{\beta}{\text{sinh}\beta}\right)\right], \hspace{0.2in} 
	\label{Arruda Boyce 8 chain free energy density function}
\end{equation}
where 
\begin{equation}
	\mu>0 \hspace{0.1in} \text{and} \hspace{0.1in} \overline{\lambda}_L > 0,
    \label{Positive value constraint on mu and lambdaL}
\end{equation}
with $\mu$ representing the ground state shear modulus, $\overline{\lambda}_L$ being the network locking stretch parameter accounting for finite chain extensibility, $\beta = \mathcal{L}^{-1}\left(\dfrac{\overline{\lambda}}{\overline{\lambda}_L}\right)$ and $\mathcal{L}(.) = \text{coth}(.) - \dfrac{1}{(.)}$ denoting the Langevin function. 

However, with new subchains being formed in their stress-free, relaxed configurations, we need to reformulate the free energy density function in \eqref{Arruda Boyce 8 chain free energy density function}. With the reasonable assumption of the reaction rate parameters being approximately constant from Section \ref{Force dependence of reaction rate parameters}, we make a few additional assumptions that enable the following analysis : (i) We only account for subchains between two B:O coordinate-bond dynamic crosslinks or two entanglement lockups and assume decoupled first order kinetics for these two types of subchains. (ii) We start from a spatially homogenous, stress-free, equilibrium state of the kinetics at $t$=0, i.e., $n_s(0)$ and $n_{ns}(0)$ related through $n_{ns}(0) = \bigl[k_{ns}/k_s\bigr] n_s(0)$. As a result, $n_s$, $n_{ns}$ will be spatially homogenous and constant with values equal to $n_s(0)$, $n_{ns}(0)$ respectively. The subchains formed at $t^{'}$ ($0 < t^{'} < t$) and surviving till the current time $t$ are stretched with respect to their stress-free, relaxed configurations corresponding to $t^{'}$. This can be quantitatively denoted by $\textbf{F}(t/t^{'})$, the relative deformation gradient. $\textbf{F}(t/t^{'}) = \textbf{F}(t) \textbf{F}^{-1}(t^{'})$ is a temporal decomposition of $\textbf{F}$ which maps the deformed configuration at $t^{'}$ to the deformed configuration at $t$. Hence, it is necessary to keep track of the subchain formation times. The development of the following expression tracking the subchain formation times, which holds individually for both types of subchains, is shown in Appendix A.1. 

\begin{equation}
    n_s(t) = \underbrace{n_s(0) e^{-k_{ns}t}}_\text{$n_s(t/0)$} + \int_{0}^{t} \underbrace{n_{ns} (0) k_s e^{-k_{ns} (t-t^{'})}}_\text{$n_s(t/t^{'})$} dt^{'}.
    \label{ns evolution integral 1}
\end{equation}

The first term on the right-hand side of \eqref{ns evolution integral 1} denoted by $n_s(t/0)$ corresponds to subchains surviving from $t$ = 0 till current time $t$, and the integrand in the second term denoted by $n_s(t/t^{'})$ corresponds to the newly formed subchains at $t^{'} \: (0 < t^{'} < t)$ and surviving till $t$. The stretch levels of $n_s(t/0)$ and $ n_s(t/t^{'})$ are quantified with $\overline{\lambda}(t/0)$ and $ \overline{\lambda}(t/t^{'})$ respectively. We use the notations $n_s$, $n_{ns}$ for $n_{s}(0)$, $n_{ns}(0)$ respectively in the following discussion for conciseness. 
 With

\begin{equation}
    \mu^{bd} = \mu^{bd}(0) = n^{bd}_s \text{k}_\text{B} \theta, \:\: \mu^{el} = \mu^{el}(0) = n^{el}_s \text{k}_\text{B} \theta,
    \label{Ground state shear modulii, B:O dative and physical crosslinks}
\end{equation}

denoting the ground state shear modulus corresponding to a network with only B:O coordinate-bond dynamic crosslinks and the ground state shear modulus corresponding to a network with only entanglement lockups respectively, $\psi^{(B1)}$ following \eqref{Arruda Boyce 8 chain free energy density function} and \eqref{ns evolution integral 1} is given as

\begin{equation}
\begin{split}
     \psi^{(B1)} &= \underbrace{\Bigl(\dfrac{n^{bd}_s(t/0)}{n^{bd}_s}\Bigr) \mu^{bd} \Bigl[\overline{\lambda}^{bd}_L\Bigr]^{2} \: \zeta^{bd}(t/0)}_\text{For subchains surviving from $t$=0} + \underbrace{\int_{0}^{t} \Bigl(\dfrac{n^{bd}_s(t/t^{'})}{n^{bd}_s}\Bigr) \mu^{bd} \Bigl[\overline{\lambda}^{bd}_L\Bigr]^2 \: \zeta^{bd}(t/t^{'}) \: dt^{'}}_\text{For subchains formed at $t^{'}$}, \\ 
     & = e^{-k^{bd}_{ns} t}  \mu^{bd}  \Bigl[\overline{\lambda}^{bd}_L\Bigr]^{2} \: \zeta^{bd}(t/0) + \int_{0}^{t}  \dfrac{n^{bd}_{ns}}{n^{bd}_s}  k^{bd}_s e^{-k^{bd}_{ns}(t-t^{'})} \mu^{bd} \Bigl[\overline{\lambda}^{bd}_L\Bigr]^2 \: \zeta^{bd}(t/t^{'}) \: dt^{'},
\end{split}
\label{Contribution to free energy, B:O dative subchains (Mechanism B)}     
\end{equation}
 
 where 
 
 \begin{equation}
\zeta^{bd}(t/t^{'}) = \left[ \left(\dfrac{\overline{\lambda}(t/t^{'})}{\overline{\lambda}^{bd}_L}\right)\beta^{bd}(t/t^{'}) + ln\left(\dfrac{\beta^{bd}(t/t^{'})}{\text{sinh}\beta^{bd}(t/t^{'})}\right)\right].  
\label{zeta definition}
 \end{equation}

The first term on the right-hand side of the free energy equation \eqref{Contribution to free energy, B:O dative subchains (Mechanism B)} corresponds to the free energy contribution from subchains surviving from $t$ = 0 till current time $t$, and the integrand in the second term corresponds to free energy contribution from the newly formed subchains at $t^{'}$ and surviving till $t$. The Cauchy stress follows from standard calculations. $\textbf{T}^{(B1)}$ corresponding to $\psi^{(B1)}$ is given as 

\begin{equation}
\textbf{T}^{(B1)} =  e^{-k^{bd}_{ns} t} \mu^{bd} \boldsymbol{\Gamma}^{bd}(t/0) +  \int_{0}^{t} \dfrac{n^{bd}_{ns}}{n^{bd}_s} k^{bd}_s e^{-k^{bd}_{ns} (t-t^{'})} \mu^{bd} \: \boldsymbol{\Gamma}^{bd}(t/t^{'}) \: dt^{'},
\label{Contribution to T(B), B:O subchains (Mechanism B)}
\end{equation}

where 

\begin{equation}
    \boldsymbol{\Gamma}^{bd}(t/t^{'}) = J^{-1}(t/t^{'}) \dfrac{\overline{\lambda}^{bd}_L}{\overline{\lambda}(t/t^{'})} \mathcal{L}^{-1}\Bigl[\dfrac{\overline{\lambda}(t/t^{'})}{\overline{\lambda}^{bd}_L}\Bigr] \Bigl(\mathbf{B}_{dis}(t/t^{'})\Bigr)_0,
    \label{TB common terms convention definition}
\end{equation}

and

\begin{equation}
	\mathbf{B}_{dis}= \mathbf{F}_{dis} (\mathbf{F}_{dis})^\top=J^{-2/3} \mathbf{B}
    \label{Distortional part of B definition}
\end{equation}

denotes the distortional part of the left Cauchy Green deformation tensor $\mathbf{B}$ with $\mathbf{F}_{dis}$ as defined in \eqref{Distortional part of F and C definition}. \\
 
The twists and knots which result in temporary entanglement lockups at high strain rates are expected to have a similar mechanical effect as the B:O coordinate-bond dynamic crosslinks. Hence, $\psi^{(B2)}$ is similarly given as

\vspace{-0.5cm}
  
 \begin{equation}
     \psi^{(B2)} = e^{-k^{el}_{ns} t}  \mu^{el}  \Bigl[\overline{\lambda}^{el}_L\Bigr]^{2} \: \zeta^{el}(t/0) + \int_{0}^{t} \dfrac{n^{el}_{ns}}{n^{el}_s} k^{el}_s e^{-k^{el}_{ns}(t-t^{'})} \mu^{el} \Bigr[\overline{\lambda}^{el}_L\Bigr]^2 \: \zeta^{el}(t/t^{'}) \: dt^{'},
     \label{Contribution to free energy, physical subchains (Mechanism B)}
 \end{equation}

with 

 \begin{equation}
\zeta^{el}(t/t^{'}) = \left[ \left(\dfrac{\overline{\lambda}(t/t^{'})}{\overline{\lambda}^{el}_L}\right)\beta^{el}(t/t^{'}) + ln\left(\dfrac{\beta^{el}(t/t^{'})}{\text{sinh}\beta^{el}(t/t^{'})}\right)\right].  
\label{zeta definition er}
 \end{equation}

 $\textbf{T}^{(B2)}$ corresponding to $\psi^{(B2)}$ is 
\vspace{-0.5cm}

\begin{equation}
\textbf{T}^{(B2)} =  e^{-k^{el}_{ns} t} \mu^{el} \boldsymbol{\Gamma}^{el}(t/0) +  \int_{0}^{t} \dfrac{n^{el}_{ns}}{n^{el}_s} k^{el}_s e^{-k^{el}_{ns} (t-t^{'})} \mu^{el} \: \boldsymbol{\Gamma}^{el}(t/t^{'}) \: dt^{'},
\label{Contribution to T(B), physical subchains (Mechanism B)}
\end{equation}

where 

\begin{equation}
    \boldsymbol{\Gamma}^{el}(t/t^{'}) = J^{-1}(t/t^{'}) \dfrac{\overline{\lambda}^{el}_L}{\overline{\lambda}(t/t^{'})} \mathcal{L}^{-1}\Bigl[\dfrac{\overline{\lambda}(t/t^{'})}{\overline{\lambda}^{el}_L}\Bigr] \Bigl(\mathbf{B}_{dis}(t/t^{'})\Bigr)_0.
    \label{TB common terms convention definition 2}
\end{equation}

\vspace{0.05in}
\subsubsection{Numerical update procedure for $\textbf{T}^{(B1)}$ and $\textbf{T}^{(B2)}$}
\vspace{0.05in}

Numerical implementation of the constitutive model is important for simulating complex, three-dimensional problems. We focused on the finite element program ABAQUS that allows the implementation of new constitutive models through user material subroutines. However, only the deformation gradients from the previous and current time steps are returned by ABAQUS at each material point. Numerical implementation of the analytical expressions for $\textbf{T}^{(B1)}$ in \eqref{Contribution to T(B), B:O subchains (Mechanism B)} and $\textbf{T}^{(B2)}$ in \eqref{Contribution to T(B), physical subchains (Mechanism B)} will require the storage of the entire deformation gradient history $\bigl($the integrands in \eqref{Contribution to T(B), B:O subchains (Mechanism B)}, \eqref{Contribution to T(B), physical subchains (Mechanism B)} involve $\boldsymbol{\Gamma}(t/t^{'}): 0<t^{'}<t$, a function of the relative deformation gradient $\textbf{F}(t/t^{'})$$\bigr)$ along with many state-dependent variables at each material point. The resulting storage requirement will be huge, making numerical implementation extremely challenging. We outline an approximate numerical update procedure for $\textbf{T}^{(B1)}$ and $\textbf{T}^{(B2)}$, which requires deformation gradients only from the previous and current time steps at each material point in Appendix A.2 for brevity. 

\vspace{0.05in}
\section{Material parameters in the model}
\label{Calibration}
\vspace{0.05in}
\subsection{Calibration}
\vspace{0.05in}

The set of model parameters in the constitutive model to be calibrated are

\begin{equation}
    \{k^{bd}_{s}, k^{bd}_{ns}, n^{bd}_{s}, k^{el}_{s}, k^{el}_{ns}, n^{el}_{s}, G, K, \nu_0, m, \alpha_p, S_0, h, \overline{\lambda}^{bd}_L, \overline{\lambda}^{el}_L\}
    \label{Calibration parameters}
\end{equation}

$k^{bd}_{ns}$ and $n^{bd}_s$ were obtained in Section \ref{B:O dative dynamic crosslinks} using the shear stress relaxation response. Following the conclusion in Section 3.1.1, we choose $k^{bd}_s \approx k^{bd}_{ns}$. $n^{el}_s$ was estimated from the ground state shear modulus $\mu^{el}$ needed to achieve the stress magnitudes at 4500 s$^{-1}$ and 8500 s$^{-1}$. $\overline{\lambda}^{bd}_L$ and $\overline{\lambda}^{el}_L$ were evaluated using the calibrated model parameters $n^{bd}_{s}$, $n^{el}_{s}$ and standard physical arguments discussed in Section {\ref{Estimates}}. We assume $k^{el}_{s} \approx k^{el}_{ns}$ and poisson's ratio relating $G$ and $K$ equal to 0.49. The list in {\eqref{Calibration parameters}} then reduces to 

\begin{equation}
    \{k^{el}_{ns}, G, \nu_0, m, \alpha_p, S_0, h\}.
    \label{Calibration parameters 2}
\end{equation}

A one-dimensional version of the constitutive model was numerically implemented in MATLAB for further calibration. $k^{el}_{ns}$ was estimated such that the stress contributions from mechanism $B2$ using $n^{el}_s$ and $\overline{\lambda}^{el}_L$ evaluated earlier are insignificant in the strain rate range of $10^{-3}$ s$^{-1}$ - $10^{0}$ s$^{-1}$, where mechanisms $A$ and $B1$ are the dominant mechanisms. Mechanism $A$ provides the primary contribution to the mechanical response at very slow strain rates. Initial estimates for the remaining parameters $G, \nu_0, m, \alpha_p, S_0, h$ from (4.2) related to Mechanism $A$ were obtained by fitting the stress contribution from Mechanism $A$ to the experimental stress-strain curve at 10$^{-3}$ s$^{-1}$ using the lsqcurvefit function (nonlinear curve-fitting in a least-squares sense) with the trust-region-reflective algorithm in MATLAB. The obtained initial estimates for the model parameters in (4.2), which may not be unique, were slightly tuned to best capture all the experimental stress-strain curves from 10$^{-3}$ s$^{-1}$ to 8500 s$^{-1}$. The values of model parameters\footnote{The one-dimensional parameters (consider 1D uniaxial stress and strain are represented with $\sigma$ and $\epsilon$, respectively) obtained from the calibration procedure except for the list of parameters
$\{\nu_{0},\alpha_p,S_0,h\}$ remain the same when used in the three-dimensional equations of the constitutive model presented. Noting that $\tau \nu = \sigma \dot{\epsilon}, \sigma = \sqrt{3}{\tau}, \dot{\epsilon} = \dfrac{\nu}{\sqrt{3}}$, the parameters $\{\nu_{0},\alpha_p,S_0,h\}$ can be converted from their one-dimensional form to their three-dimensional shear form using the following conversion relations: $\nu_0^{(shear)} = \sqrt{3} \nu_0^{(comp)}, \alpha_p^{(shear)} = \dfrac{1}{\sqrt{3}} \alpha_p^{(comp)}, S_0^{(shear)} = \dfrac{1}{\sqrt{3}} S_0^{(comp)}, h^{(shear)} = \dfrac{1}{{3}} h^{(comp)}$.} after calibration are listed in Table \ref{Material parameters}.

\begin{table}[!h]
\begin{center}
\begin{tabular}{ | m{2.5cm} | m{2 cm} | } 
    
  \hline
  Parameter & PBS\\
  \hline

  $k^{bd}_{s}$ (s$^{-1}$) & 0.35 \\

  $k^{bd}_{ns}$ (s$^{-1}$) & 0.35 \\

  $n^{bd}_s$ (m$^{-3}$) &  6.4 x 10$^{24}$\\

  $k^{el}_{s}$ (s$^{-1}$) & 2 x 10$^{3}$ \\
   
  $k^{el}_{ns}$ (s$^{-1}$) & 2 x 10$^{3}$ \\

  $n^{el}_s$ (m$^{-3}$) &  2 x 10$^{26}$\\
  
  \hline

   $G$ (MPa) & 0.4\\
   
   $K$ (MPa) & 20\\ 
  
   \hline

  $\nu_0$ (s$^{-1}$) & 2 x 10$^{-3}$\\
   
  $m$ (-) & 0.95\\

  $\alpha_p$ (-) & 0.11\\

  $S_0$ (kPa) & 0.6 \\
  
  $h$ (kPa) & 37.7\\
  
  \hline

  \vspace{0.1cm}
 
  $\overline{\lambda}^{bd}_L$ (-) & 37.4\\

  $\overline{\lambda}^{el}_L$ (-) & 6.7\\

  \hline

  $\rho$ (kg m$^{-3}$) & 1100 \\

  \hline
\end{tabular}
\caption{\label{Material parameters} Model parameters for PBS.}
\end{center}
\end{table}

\begin{figure}[h!]
    \begin{center}
		\includegraphics[width=\textwidth]{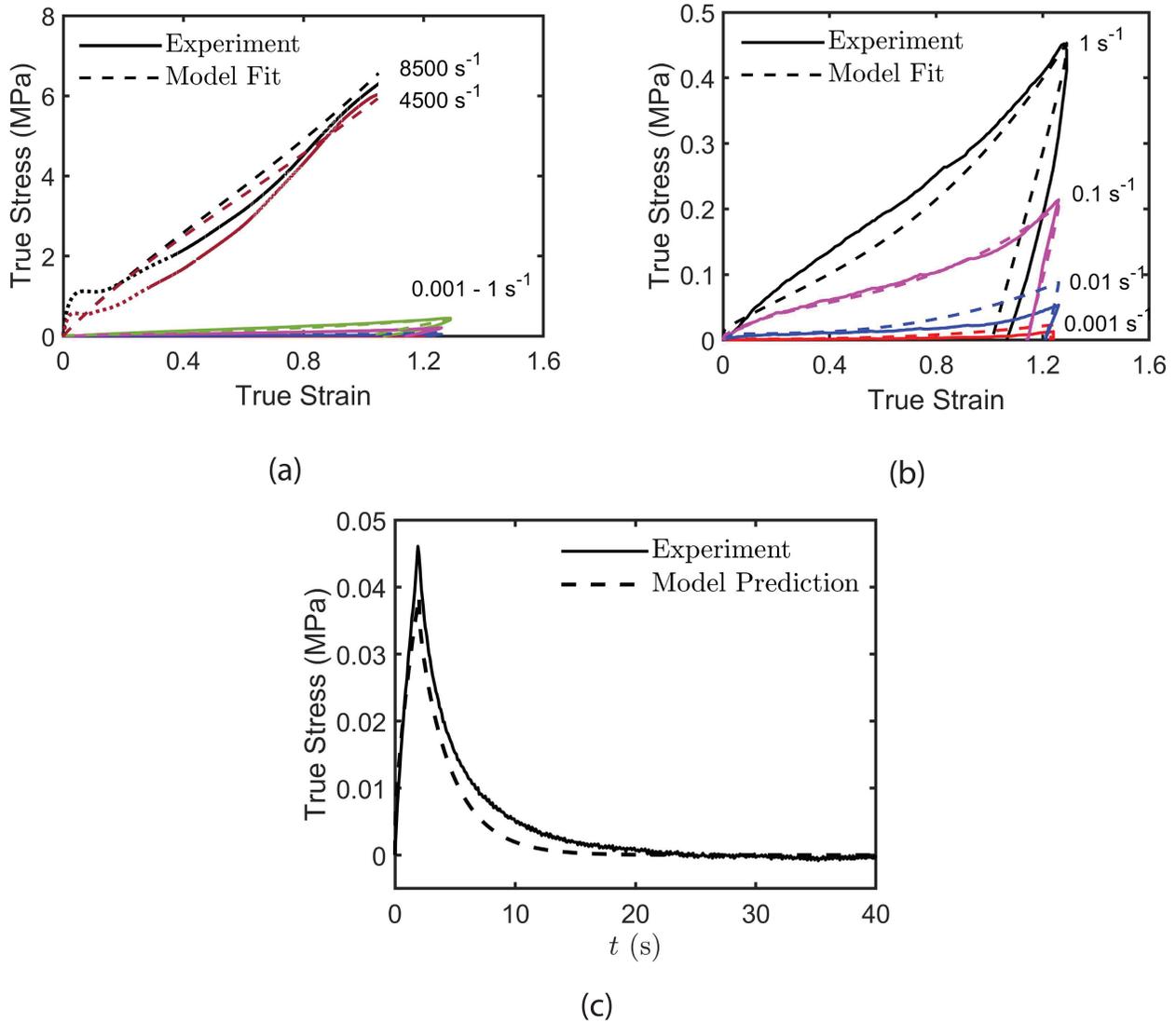}
	\end{center}
    \caption{{\small(a) Fit of the model (dashed lines) to experimental compression stress-strain curves (solid lines) for PBS over strain rates from 0.001 s$^{-1}$ to 8500 s$^{-1}$. The dotted lines in the experimental curves for 4500 s$^{-1}$ and 8500 s$^{-1}$ indicate the regions where dynamic force equilibrium is not achieved. (b) Zoomed-in view of the fit to the experimental compression response for relatively lower strain rates from 0.001 s$^{-1}$ to 1 s$^{-1}$. (c) Comparison of the model predicted large strain compression stress relaxation response (compressed at a strain rate of 0.1 s$^{-1}$ to 20$\%$ true strain, then the strain was held constant) against the experimental result. The good agreement suggests that B:O coordinate-bond dynamic crosslink relaxation is the dominant relaxation mechanism in the quasi-static (10$^{-3}$ s$^{-1}$) to moderate (10$^{0}$ s$^{-1}$) strain rate range even at large strains.}}
    \label{Model fit and large strain relaxation}
\end{figure}

The fit of the constitutive model to the experimental compression stress-strain curves for PBS over 0.001 s$^{-1}$ to 8500 s$^{-1}$ strain rate is shown in Figure \ref{Model fit and large strain relaxation}. The constitutive model reasonably fits the non-linear, large strain, loading-unloading, and substantial rate-dependent stiffening observed in PBS. In Section \ref{B:O dative dynamic crosslinks}, the small strain shear stress relaxation and oscillatory shear response supported the assumption of B:O coordinate-bond dynamic crosslink relaxation being the dominant relaxation mechanism in the quasi-static (10$^{-3}$ s$^{-1}$) to moderate (10$^{0}$ s$^{-1}$) strain rate range. The model predicts the stress relaxation response for a large strain case  (compressively loaded at a strain rate of 0.1 s$^{-1}$ to 20\% true strain, and then the strain was held constant) well as shown in Figure \ref{Model fit and large strain relaxation}(c). This further indicates that the B:O coordinate-bond dynamic crosslink relaxation is the dominant relaxation mechanism for slow to moderate rate loadings for both small and large strains.  

\vspace{0.1in}
\subsection{Estimates of subchain lengths in PBS}
\label{Estimates}
 \vspace{0.05in}
 
 With the average number of monomers in the subchains as $N_s$ and monomer length as $b$, the root mean squared (rms) end-to-end distance is given as $\overrightarrow{r_s} = \sqrt{<\textbf{\emph{r}}_s^2>} = \sqrt{{N}_s}b$. Length of the dimethylsiloxane monomer, $b$ can be estimated using the Si-O bond length ($\approx 0.164 \: \text{nm}$) and O-Si-O bond angle ($\approx 110^{\text{o}}$) as $b$ $\approx$ 2 x 0.164 x sin(55$^{\text{o}}$) nm, i.e., $b \approx$ 0.27 nm (\citep{SiObondlengthangle}). Estimates of the average molecular mass of subchains can be attained using $ {M}_{s} = \rho \text{R} / (\text{k}_{\text{B}} n_s)$, where R is the gas constant. $M^{bd}_s$ and $M^{el}_s$ corresponding to B:O coordinate-bond dynamic crosslinks and entanglement lockups evaluated using model parameters $n^{bd}_s$, $n^{el}_s$ and the earlier relation are approximately 103.55 kg/mol and 3.3 kg/mol, respectively. Molecular mass ($\approx$ 0.074 kg/mol) and length of the dimethylsiloxane monomer, $b$, can be used to approximate the average subchain contour lengths, $L^{bd}_s$ and $L^{el}_s$. Scaling the dimethylsiloxane monomer length with the ratio of $M_s$ and dimethylsiloxane monomer molecular mass results in \emph{$L^{bd}_s$ $\approx$ 378 nm, $L^{el}_s$ $\approx$ 12 nm} and $N^{bd}_s \approx 1400$, $N^{el}_s \approx 45$.\footnote{Note that the network locking stretch parameters $\overline{\lambda}^{bd}_L$ and $\overline{\lambda}^{el}_L$ can be calculated using the physical arguments that the stretch locking parameter is the ratio of the fully extended subchain length ($\approx {N}_s b$) to the corresponding initial root mean squared length ($\overrightarrow{r_s} = \sqrt{{N}_s}b$). Therefore, $\overline{\lambda}^{bd}_L \approx \sqrt{{N}^{bd}_s} \approx 37.4$ and $\overline{\lambda}^{el}_L \approx \sqrt{{N}^{el}_s} \approx 6.7$.} Using $\overrightarrow{r_s} = \sqrt{{N}_s}b$, we get rms end-to-end distance for subchains between B:O coordinate-bond dynamic crosslinks to be \emph{$\overrightarrow{r_s}^{bd} \approx$ 10 nm}, and for the subchains between entanglement lockups as \emph{$\overrightarrow{r_s}^{el} \approx$ 1.8 nm}. 

\vspace{0.1in}
\section{Validation experiments and numerical simulations}
\label{Validation}

 We conducted a few validation experiments not used to determine the parameters of our model and compared the experimental results against those from the corresponding three-dimensional numerical simulations. Indentation experiments resulting in inhomogeneous deformations were carried out on PBS rectangular slabs (blocks). Two distinct types of indenters were considered, (i) a conical indenter causing local deformation at the center of the block and (ii) a long, triangular prismatic indenter deforming blocks at two different loading rates. The experiments were conducted using UniVert Mechanical Test System (CellScale). The bottom surfaces of the blocks were fixed using a cyanoacrylate adhesive. The total external force required along the direction of indentation for a prescribed indenter displacement profile was measured in each case. The indenter noses were designed to have tip radii of 2 mm to prevent the tearing of PBS. The constitutive model was numerically implemented (explicit implementation) in ABAQUS/Explicit (2018)\footnote{Explicit implementation procedure was chosen considering the applications with relatively short timescales of loadings for PBS. ABAQUS/Explicit uses the Green-Naghdi objective stress rate and the symmetric part of the velocity gradient as the objective deformation rate when using VUMAT for the finite deformation behavior of a material.} by writing a user-defined material subroutine (VUMAT). We have developed an approximate numerical method that allows for stress updates using only deformation gradients at the previous and current time steps. This eliminates the need for storage of the entire deformation gradient history at each material point and the associated large data storage requirements. The mathematical development is presented in Appendix A.2. The finite element software ABAQUS, has been applied for a variety of engineering simulations to solve problems involving complex geometry, loading, and material behavior (\citep{Srivastava2011STRESSPIPE, Bai2021AEngineering,  Zhong2021AGrowth, Niu2022SimulationMeasurements, Niu2022engwithcomp}). The standard implementation procedure in ABAQUS/Explicit using VUMAT has been followed. It can be summarized as using the deformation gradient for the current time step provided by ABAQUS at each material point to perform explicit integration of the differential equations involved. This allows for the straightforward calculation of the total Cauchy stress using (3.14), (3.16), (A.26), (A.27), and an update of the state variables. Polylactic acid indenters were modeled as rigid analytical surfaces. Contact between the indenters and PBS blocks was assumed to be frictionless as the interfaces were lubricated during the experiments. The dimensions of the blocks used were 20$\pm$2 mm x 20$\pm$2 mm x 8$\pm$1 mm. 

\vspace{0.1in}
\subsection{Conical indenter}
\label{Conical Indenter}

\begin{figure}[!]
    \begin{center}
		\includegraphics[width=12cm]{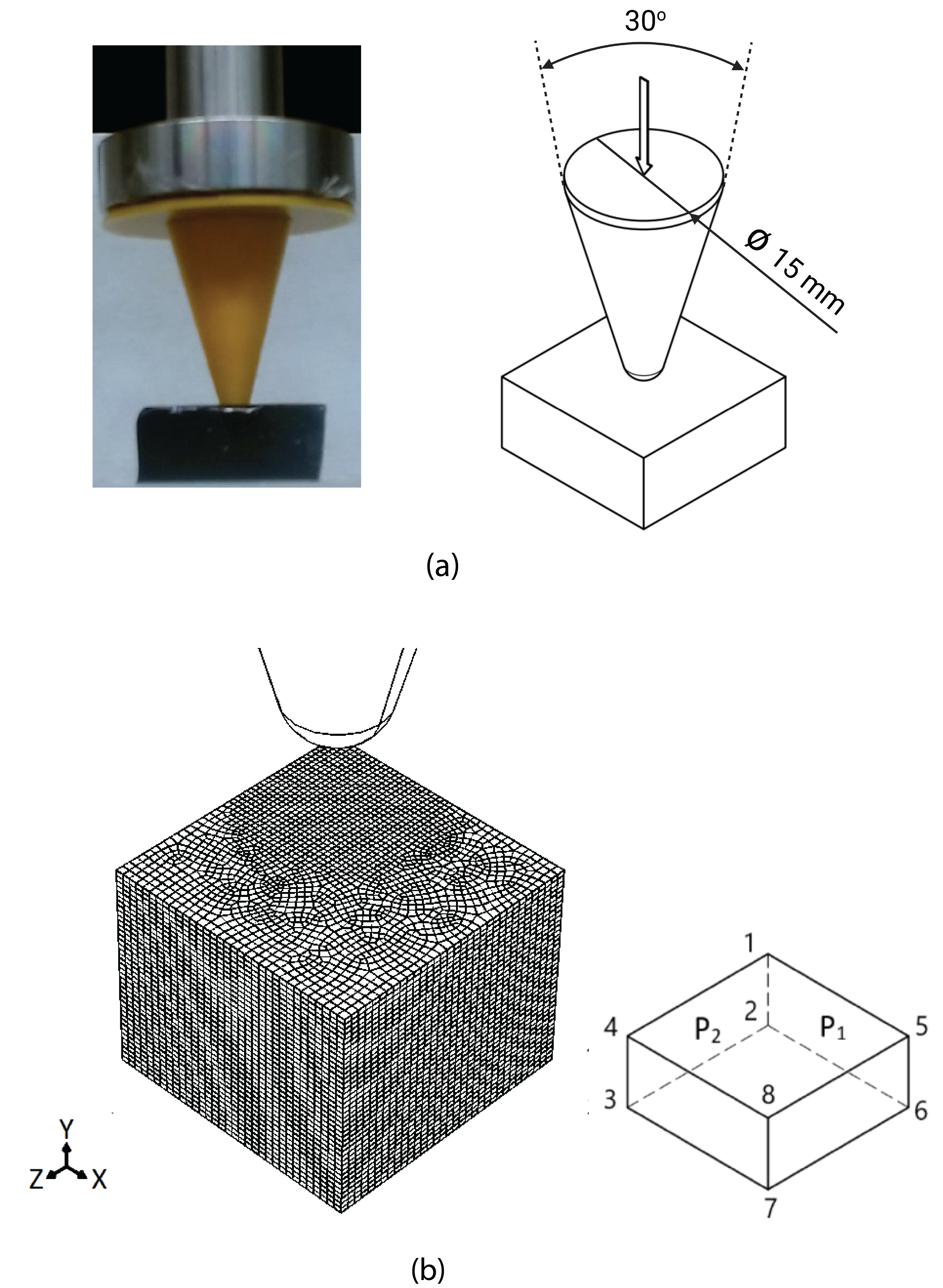}
	\end{center}
    \caption{{\small(a) Schematic and dimensions in mm for the conical indenter experiment. The PBS block was colored for easy visualization. (b) The experiment is symmetric about the planes denoted by $\text{P}_1$, $\text{P}_2$, and as a result, only a quarter block needs to be considered with appropriate boundary conditions. Finite element mesh used for the quarter block simulation, the conical indenter is modeled as a rigid surface. The bottom surface 2-6-7-3 of the quarter block is fixed while the surfaces 1-2-3-4 on $\text{P}_2$ and 1-5-6-2 on $\text{P}_1$ are prescribed the boundary conditions $u_{\text{X}}=u^{\text{R}}_{\text{Y}}=u^{\text{R}}_{\text{Z}}=0$ and $u_{\text{Z}}=u^{\text{R}}_{\text{X}}=u^{\text{R}}_{\text{Y}}=0$, respectively.}}
    \label{10 Drawing mesh and schematic}
\end{figure}

\begin{figure}[!]
    \begin{center}
		\includegraphics[width=\textwidth]{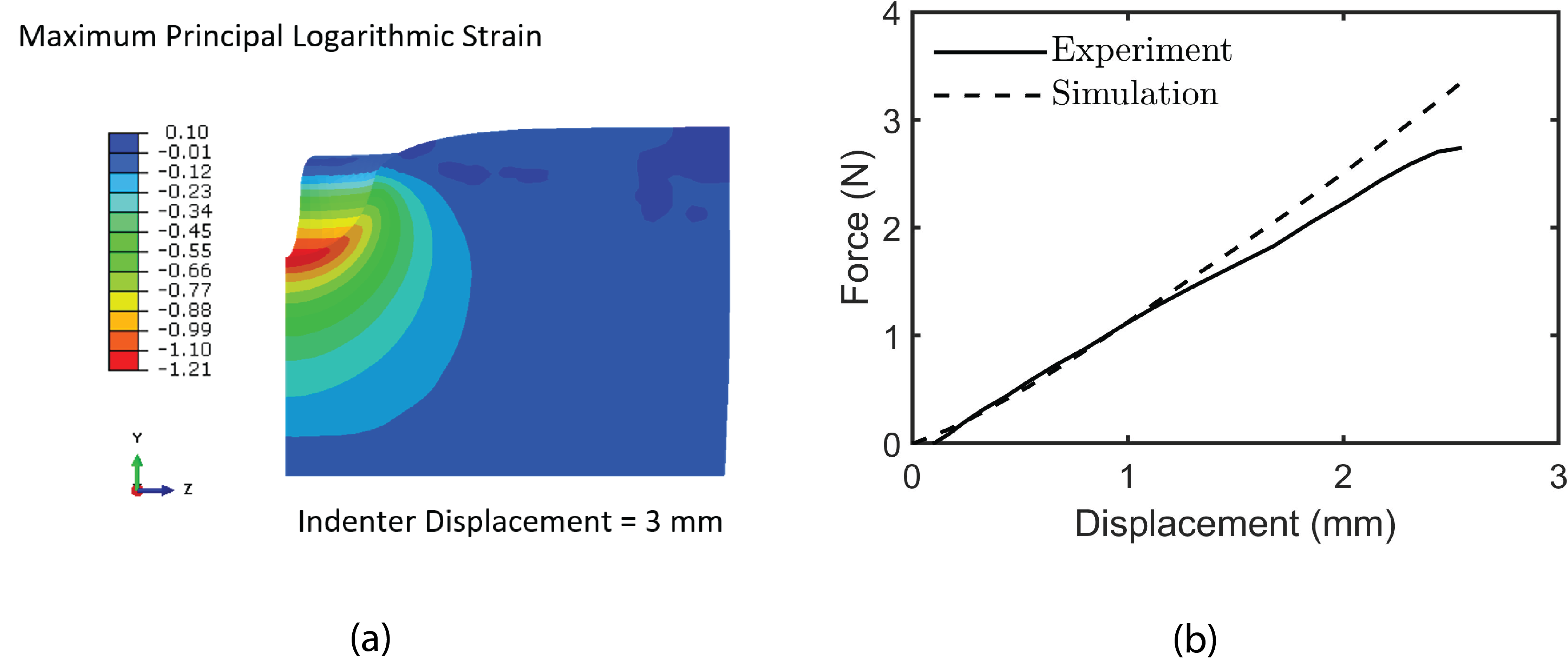}
	\end{center}
    \caption{{\small(a) Contours of maximum principal logarithmic strain on the YZ plane of the deformed quarter PBS block. (b) Comparison of the numerically predicted indentation force with the corresponding experimental result.}}
    \label{10 Results}
\end{figure}

The conical indenter was prescribed a linear displacement ramp to 3 mm displacement with a speed of 10 mm/s.  Figure \ref{10 Drawing mesh and schematic}(a) displays the experiment schematic and dimensions in mm of the conical indenter. Considering the symmetries of the experiment about the planes $\text{P}_1$ and $\text{P}_2$ in Figure \ref{10 Drawing mesh and schematic}(b), only a quarter of the block needs to be simulated with appropriate boundary conditions. The quarter block was meshed using 76960 ABAQUS C3D8R three-dimensional, reduced integration stress elements as shown in Figure \ref{10 Drawing mesh and schematic}(b). The indenter initially touching the vertex 1 of the representative numbered quarter block in Figure \ref{10 Drawing mesh and schematic}(b) moves along -\textbf{e}$_{\text{Y}}$ direction. The bottom surface 2-6-7-3 is fixed while the surfaces 1-2-3-4 on $\text{P}_2$ and 1-5-6-2 on $\text{P}_1$ are prescribed the boundary conditions $u_{\text{X}}=u^{\text{R}}_{\text{Y}}=u^{\text{R}}_{\text{Z}}=0$ and $u_{\text{Z}}=u^{\text{R}}_{\text{X}}=u^{\text{R}}_{\text{Y}}=0$ respectively accounting for the symmetries discussed earlier. $u$, $u^{\text{R}}$ denote the translational and rotational degrees of freedom of a point about the axis indicated in the subscript. Figure \ref{10 Results}(a) shows the contours of maximum principal logarithmic strain on the YZ projection of the fully deformed quarter block showing the complex three-dimensional, inhomogeneous deformations involved. The maximum local strain rate in the specimen (from the simulation) during the loading is $\sim$10 s$^{-1}$. Numerical prediction for the indentation force agrees well with the experimental result, as seen in Figure \ref{10 Results}(b).

\vspace{0.1in}
\subsection{Triangular prismatic indenter}
 Two linear ramp displacement profiles to 3 mm at speeds of (i) 0.7 mm/s and (ii) 2.5 mm/s were prescribed to a triangular prismatic indenter. Figure \ref{5,15 Drawing mesh and schematic}(b) shows the schematic for the experiment and dimensions for the indenter in mm. The experiment is symmetric about the planes $\text{P}_1$ and $\text{P}_2$ in Figure \ref{5,15 Drawing mesh and schematic}(c), and hence only a quarter of the block needs to be simulated with appropriate boundary conditions. The quarter block was uniformly meshed using 29403 (27 along the thickness) ABAQUS C3D8R three-dimensional, reduced integration stress elements as shown in Figure \ref{5,15 Drawing mesh and schematic}(c). The indenter initially touching the edge 1-4 of the representative numbered quarter block in Figure \ref{5,15 Drawing mesh and schematic}(c) moves along -\textbf{e}$_{\text{Y}}$ direction. The bottom surface 2-6-7-3 is fixed while the surfaces 1-2-3-4 on $\text{P}_2$ and 1-5-6-2 on $\text{P}_1$ are prescribed the boundary conditions $u_{\text{X}}=u^{\text{R}}_{\text{Y}}=u^{\text{R}}_{\text{Z}}=0$ and $u_{\text{Z}}=u^{\text{R}}_{\text{X}}=u^{\text{R}}_{\text{Y}}=0$, respectively. Figure \ref{5,15 Results}(a) shows the contours of maximum principal logarithmic strain on the isometric view and XY projection of the completely deformed full block for the indentation speed of 0.7 mm/s showing the inhomogeneous deformations. The maximum local strain rates in the specimens during the loadings are of the order of 0.1 s$^{-1}$ and 1 s$^{-1}$ for the indenter speeds of 0.7 mm/s and 2.5 mm/s, respectively. Good agreement between the predictions from simulations and experimental results for the indentation forces can be seen in Figure \ref{5,15 Results}(b). 

\begin{figure}[!]
    \begin{center}
		\includegraphics[width=\linewidth]{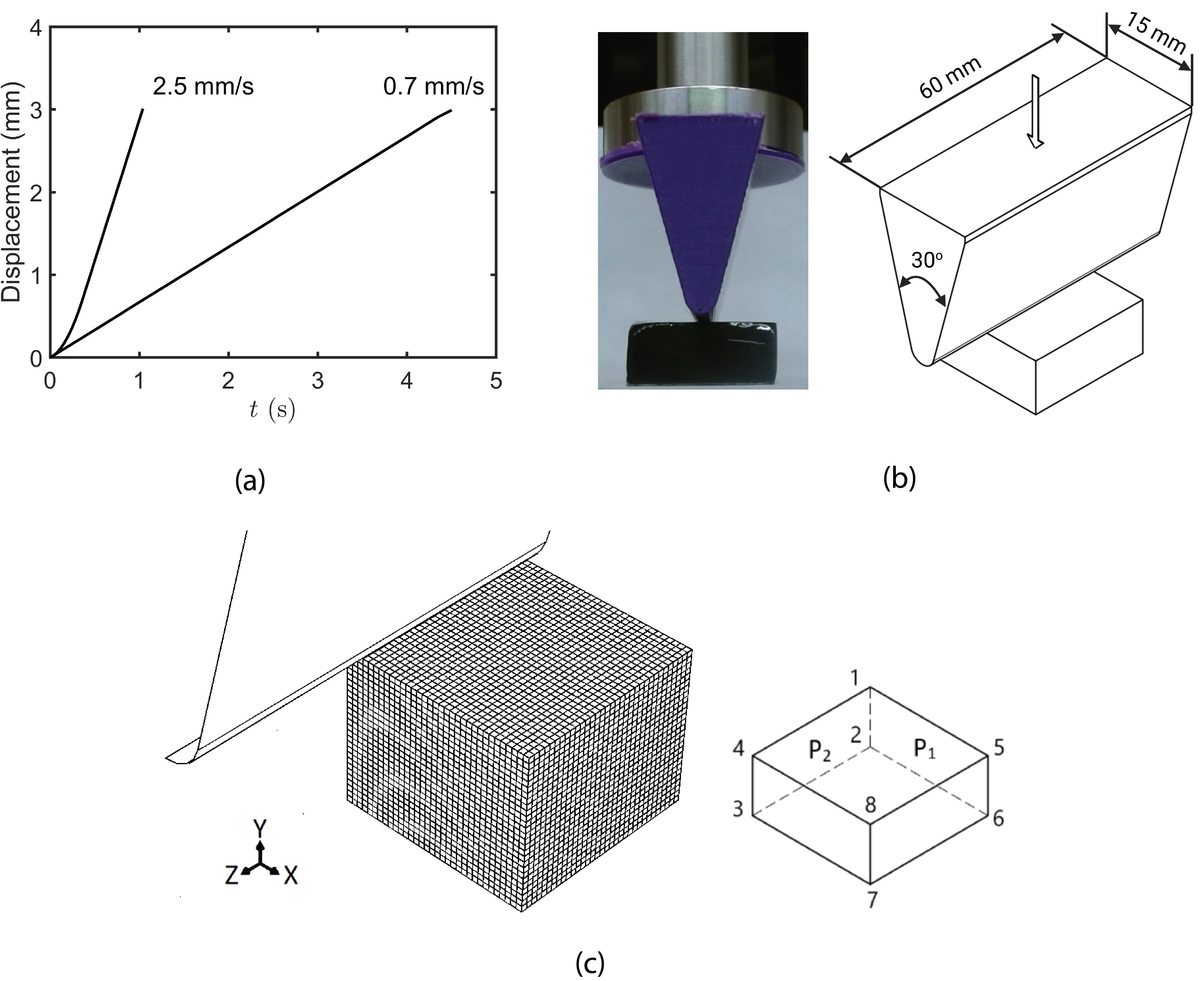}
	\end{center}
    \caption{{\small(a) Linear ramp displacement profiles of the triangular prismatic indenter. (b) Schematic for the experiment and dimensions in mm for the triangular prismatic indenter. The PBS block was colored for easy visualization. (c) The experiment is symmetric about the planes denoted by $\text{P}_1$, $\text{P}_2$, and as a result, only a quarter block needs to be considered with appropriate boundary conditions. Finite element mesh used for the quarter block simulations. The triangular prismatic indenter was modeled as a rigid surface. The bottom surface 2-6-7-3 of the quarter block is fixed while the surfaces 1-2-3-4 on $\text{P}_2$ and 1-5-6-2 on $\text{P}_1$ are prescribed the boundary conditions $u_{\text{X}}=u^{\text{R}}_{\text{Y}}=u^{\text{R}}_{\text{Z}}=0$ and $u_{\text{Z}}=u^{\text{R}}_{\text{X}}=u^{\text{R}}_{\text{Y}}=0$, respectively.}}
    \label{5,15 Drawing mesh and schematic}
\end{figure}

\begin{figure}[!]
    \begin{center}
		\includegraphics[width=\textwidth]{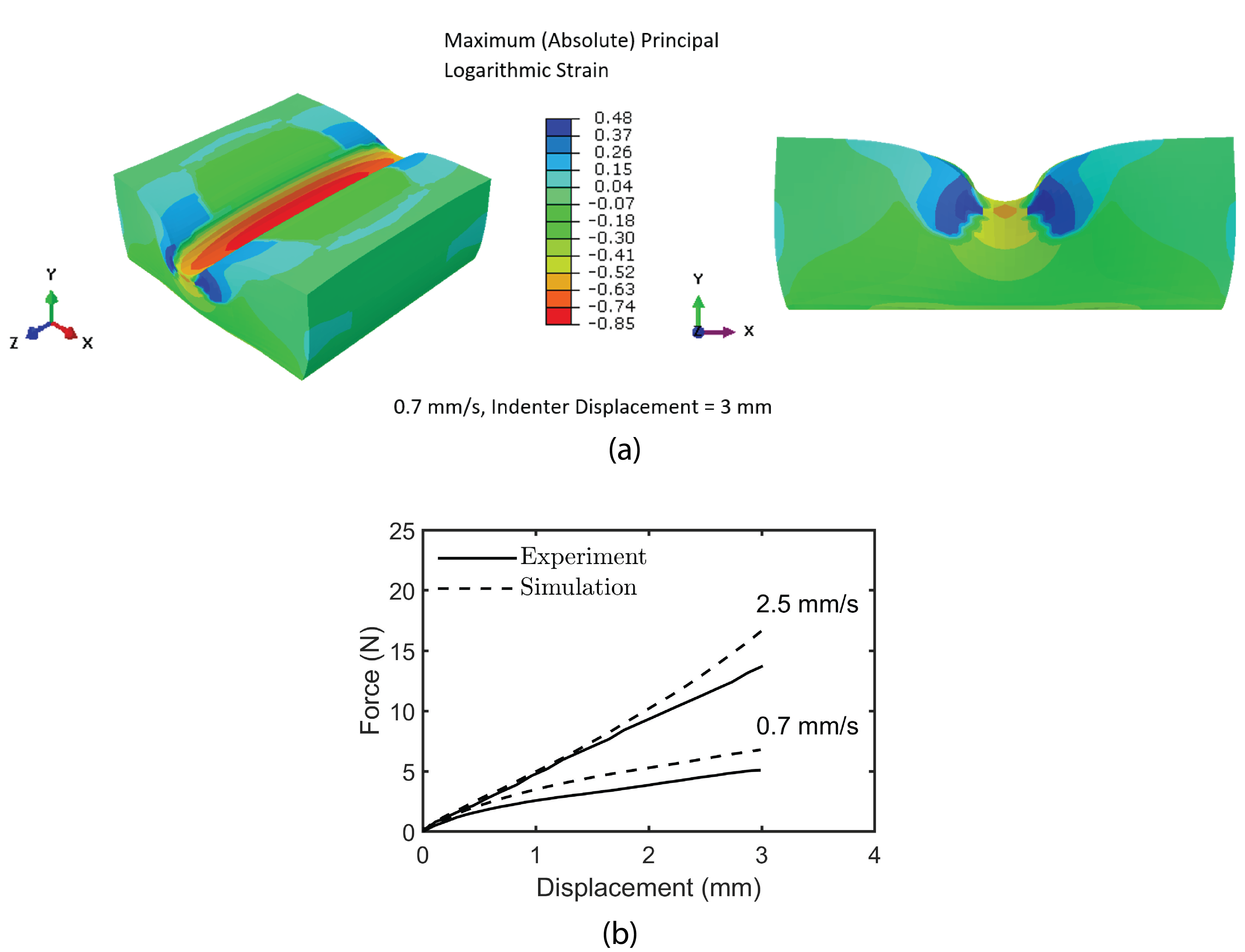}
	\end{center}
    \caption{{\small(a) Contours of maximum principal logarithmic strain in the isometric view and on the XY projection of the deformed PBS block for the indentation speed of 0.7 mm/s. (b) Comparison of the numerically predicted indentation forces for triangular prismatic indenter speeds of 0.7 mm/s and 2.5 mm/s with corresponding experimental results. The model and simulation predicted results are very close to the experimental results}}%
    \label{5,15 Results}
\end{figure}

\section{Conclusions}
\label{Closing Remarks}

We have conducted experiments and developed a finite deformation elastic-viscoplastic constitutive model for Polyborosiloxane (PBS), an important rate stiffening soft polymer with dynamic, reversible crosslinks for quasi-static (10$^{-3}$ s$^{-1}$) to high (10$^{3}$ s$^{-1}$) strain rates. Using the macroscopic observations from our experiments and the model framework, we suggest two crosslink mechanisms in PBS and their distinct macroscopic relaxation timescales ($\tau$). The dynamic network-forming reversible crosslink mechanisms are boron-oxygen coordinate-bond crosslinks ($\tau \approx$ 3 s) and temporary entanglement lockups at high strain rates acting as crosslinks ($\tau \approx$ 0.5 ms). The microstructure physics based continuum model suggests possible microstructural mechanisms and provides estimations of the subchain lengths between crosslinks from the macroscopic experiments. We estimate subchains between B:O coordinate-bond dynamic crosslinks to have root mean squared end-to-end distance \emph{$\overrightarrow{r_s}^{bd} \approx$ 10 nm}, and the subchains between entanglement lockups to have \emph{$\overrightarrow{r_s}^{el} \approx$ 1.8 nm} in PBS under undeformed state. The average subchain contour lengths are estimated to be \emph{$L^{bd}_s$ $\approx$ 378 nm} for subchains between B:O coordinate-bond dynamic crosslinks, and \emph{$L^{el}_s$ $\approx$ 12 nm} for subchains between entanglement lockups. To analyze real structures, it is essential to perform new material model based computational simulations for three-dimensional geometries. We have numerically implemented the model in the finite element software ABAQUS and have outlined a numerical update procedure to evaluate the convolution-like time integrals arising from dynamic crosslink kinetics. The approximate predictive capabilities of the constitutive model and its numerical implementation were verified by comparing the simulation results with results from validation experiments involving inhomogeneous deformations. The proposed model and its theoretical framework should extend to other soft polymers and polymer gels with dynamic, reversible crosslinks which are microstructurally similar to PBS. 

New microstructural experimental techniques in the future that enable direct measurement of dynamic crosslink formation, relaxation timescales, and associated subchain lengths and densities will allow direct evaluation of microstructural parameters and validation of some of the microstructural material parameter estimates based on the continuum-scale model. Large strain material response characterization in the intermediate strain rate regime (10$^{1}$ s$^{-1}$ - 10$^{2}$ s$^{-1}$) for soft polymers is a challenging and active area of research. In the future, new methods for accurately characterizing the stress-strain response of soft polymers in the intermediate strain rate range will be of significant benefit.

\section*{Author Contributions}
\textbf{Aditya Konale}: Investigation, Methodology, Formal analysis, Data curation, Software, Visualization, Writing – original draft preparation, review \& editing. \textbf{Zahra Ahmed}: Investigation, Writing – review \& editing. \textbf{Piyush Wanchoo}: Investigation, Writing – review \& editing. \textbf{Vikas Srivastava}: Conceptualization, Investigation, Methodology, Formal analysis, Supervision, Funding acquisition, Writing – original draft preparation, review \& editing.

\section*{Acknowledgments}
The authors thank Dr. Arun Shukla at the University of Rhode Island for guidance and for providing access to the Split Hopkinson Pressure Bar facility in the Dynamic Photo Mechanics Laboratory. The authors also thank Dr. Roberto Zenit at Brown University for providing access to the ARES G2 Rheometer. The authors thank Dr. James LeBlanc from Naval Underwater Warfare Center, Dr. Yuri Bazilevs and Dr. Pradeep Guduru at Brown University for insightful discussions on the topic. Finally, the authors would like to acknowledge the support from the Office of Naval Research (ONR), USA, to V. Srivastava under grant numbers N00014-21-1-2815 and N00014-21-1-2670.

\begin{appendices}

\section{}

\subsection{Expression for tracking subchain formation times}

We summarize the development of an expression for tracking the subchain formation times for reversible crosslinks (\citep{FMeng(2016)Stressfreeconfigmodeling1, FMeng(2016)Stressfreeconfigmodeling2}) assuming first-order kinetics (Refer to the assumptions in Section \ref{Free energy B}). After the first time step $\Delta t$, with the first order kinetics in \eqref{B:O reversible crosslink first order kinetics}, we have

\begin{equation}
    n_s(\Delta t) = n_s(0) \Bigl[1 - k_{ns}\Delta t\Bigr] + n_{ns}(0) k_s \Delta t.
    \label{ns evolution first time step 1}
\end{equation}

For $\Delta t \rightarrow$ 0, using Taylor series expansion of $e^{x}$ to simplify \eqref{ns evolution first time step 1}, we get

\begin{equation}
    n_s (\Delta t) = n_s(0) e^{-k_{ns}\Delta t} + n_{ns}(0) k_s \Delta t.
    \label{ns evolution first time step 2}
\end{equation}

Consequently, after the second time step, we have

\vspace{-0.5cm}

\begin{equation}
\begin{split}
    n_s(2\Delta t) &= n_s(0) e^{-k_{ns}\Delta t}\Bigl[1-k_{ns}\Delta t\Bigr] + n_{ns}(0) k_s \Delta t\Big[1-k_{ns} \Delta t\Bigr] + n_{ns}(0) k_s \Delta t.
\end{split}
\label{ns evolution second time step 1}
\end{equation}

Again, simplifying \eqref{ns evolution second time step 1} for small $\Delta t \rightarrow$ 0, we get

\vspace{-0.5cm}

\begin{equation}
    n_s(2\Delta t) = n_s(0) e^{-2k_{ns}\Delta t} + n_{ns}(0)k_s\Delta t e^{-k_{ns}\Delta t} + n_{ns}(0) k_s \Delta t.
    \label{ns evolution second time step 2}
\end{equation}

After N time steps, following \eqref{ns evolution first time step 2} and \eqref{ns evolution second time step 2}, $n_{s}(\text{N}\Delta t)$ can be expressed as

\vspace{-0.5cm}

\begin{equation}
    n_s(\text{N}\Delta t) = n_s(0) e^{-\text{N} k_{ns}\Delta t} + \sum_{i = 1}^{\text{N}} n_{ns}(0)k_s\Delta t e^{-(\text{N}-i)k_{ns}\Delta t}.
    \label{ns evolution Nth time step}
\end{equation}

This evolution pattern can be written concisely in an integral form for $\Delta t \rightarrow$ 0 as

\begin{equation}
    n_s(t) = n_s(0) e^{-k_{ns}t} + \int_{0}^{t} n_{ns} (0) k_s e^{-k_{ns} (t-t^{'})} dt^{'}.
    \label{ns evolution integral appendix}
\end{equation}

The first term on the right-hand side of \eqref{ns evolution integral appendix} corresponds to subchains surviving from $t$ = 0 till current time $t$, and the integrand in the second term corresponds to the newly formed subchains at $t^{'} \: (0 < t^{'} < t)$ and surviving till $t$.

\subsection{Numerical update procedure for \textbf{T}$^{(B1)}$ and \textbf{T}$^{(B2)}$}
 We briefly outline the numerical update procedure for \textbf{T}$^{(B1)}$ and \textbf{T}$^{(B2)}$ using only the deformation gradients from the previous and current time steps at each material point. Focusing on $\textbf{T}^{(B1)}$ in \eqref{Contribution to T(B), B:O subchains (Mechanism B)}, we have

\vspace{-0.5cm}

\begin{equation}
\begin{split}
    \textbf{T}^{(B1)}(t) &=  e^{-k^{bd}_{ns} t} \mu^{bd} \boldsymbol{\Gamma}^{bd}(t/0) +  \int_{0}^{t} \dfrac{n^{bd}_{ns}}{n^{bd}_s} k^{bd}_s e^{-k^{bd}_{ns} (t-t^{'})} \mu^{bd} \: \boldsymbol{\Gamma}^{bd}(t/t^{'}) \: dt^{'}.
\end{split}
\label{TB update step 2}
\end{equation}

For the next time step, i.e. $t$+$\Delta t$, \eqref{TB update step 2} can be rewritten as

\vspace{-0.5cm}

\begin{equation}
\begin{split}
    \textbf{T}^{(B1)}(t+\Delta t) &=  e^{-k^{bd}_{ns} (t + \Delta t)} \mu^{bd} \boldsymbol{\Gamma}^{bd}(t+\Delta t/0) + \int_{0}^{t+\Delta t} \dfrac{n^{bd}_{ns}}{n^{bd}_s} k^{bd}_s e^{-k^{bd}_{ns} (t+\Delta t-t^{'})} \mu^{bd} \: \boldsymbol{\Gamma}^{bd}(t+\Delta t/t^{'}) dt^{'}. 
\end{split}
\label{TB at t+delta t}
\end{equation}

Further, $\textbf{F}(t+\Delta t/t^{'})$ and $J(t+\Delta t/t^{'})$ can be split as

\vspace{-0.5cm}

\begin{equation}
    \textbf{F}(t+\Delta t/t^{'}) = \textbf{F}(t+\Delta t/t) \textbf{F}(t/t^{'}) \: \: ; \: \: J(t+\Delta t/t^{'}) = J(t+\Delta t/t) J(t/t^{'}).
    \label{Decomposition of F(t+delta t/t')}
\end{equation}

With \eqref{Decomposition of F(t+delta t/t')}, and \eqref{Distortional part of F and C definition}, $\textbf{C}_{dis}(t+\Delta t/t^{'})$ can be expressed as

\vspace{-0.5cm}

\begin{equation}
\begin{split}
    \textbf{C}_{dis}(t+\Delta t/t^{'}) &= J^{-2/3}(t/t^{'})\textbf{F}^{T}(t/t^{'}) \textbf{C}_{dis}(t+\Delta t/t) \textbf{F}(t/t^{'}).
\end{split}
\label{Decomposition of Cdis(t+delta t/t')}
\end{equation}

Using \eqref{Decomposition of Cdis(t+delta t/t')} and $\overline{\lambda} \overset{\text{def}}{=} \sqrt{\dfrac{\mathrm{tr} \mathbf{C}_{dis}}{3}}$, $\overline{\lambda}(t+\Delta t/t')$ is given as

\vspace{-0.5cm}

\begin{equation}
\begin{split}
    \overline{\lambda}(t+\Delta t/t^{'}) &= \sqrt{\dfrac{\text{tr}\Bigl[J^{-2/3}(t/t^{'})\textbf{F}^{T}(t/t^{'}) \textbf{C}_{dis}(t+\Delta t/t) \textbf{F}(t/t^{'})\Bigr]}{3}}.
\end{split}
\label{Total effective distortional stretch lambda(t+delta t/t')}
\end{equation}

For $\Delta t \rightarrow$ 0, with \eqref{Distortional part of F and C definition}, it follows that

\vspace{-0.5cm}

\begin{equation}
\begin{split}
    \textbf{C}_{dis}(t+\Delta t/t) &\rightarrow \textbf{I}.
\end{split}
\label{Distortional part of C, identity for small delta t}
\end{equation}

Under the limit of $\Delta t \rightarrow$ 0, \eqref{Total effective distortional stretch lambda(t+delta t/t')} can be approximated using \eqref{Distortional part of C, identity for small delta t}, \eqref{Distortional part of F and C definition} as

\vspace{-0.5cm}

\begin{equation}
\begin{split}
    \overline{\lambda}(t+\Delta t/t^{'}) &\rightarrow  \overline{\lambda}(t/t^{'})
    \label{lambda(t+delta t/t') = lambda(t/t') for small delta t}.
\end{split}
\end{equation}

Following the definition of the deviatoric part of a tensor, $\Bigl[\textbf{B}_{dis}(t+\Delta t/t^{'})\Bigr]_{0}$ is given as

\begin{equation}
    \Bigl[\textbf{B}_{dis}(t+\Delta t/t^{'})\Bigr]_{0} = \textbf{B}_{dis}(t+\Delta t/t^{'}) - \dfrac{1}{3} \text{tr}\Bigl[\textbf{B}_{dis}(t+\Delta t/t^{'})\Bigr].
    \label{Deviatoric part of Bdis(t+delta t/t')}
\end{equation}

$\textbf{B}_{dis}(t+\Delta t/t^{'})$ can be expressed following \eqref{Decomposition of Cdis(t+delta t/t')} as

\begin{equation}
    \textbf{B}_{dis}(t+\Delta t/t^{'}) = J^{-2/3}(t+\Delta t/t)\textbf{F}(t+\Delta t/t) \textbf{B}_{dis}(t/t^{'}) \textbf{F}^{T}(t+\Delta t/t).
    \label{Decomposition of Bdis(t+delta t/t')}
\end{equation}

Further, using \eqref{Decomposition of Bdis(t+delta t/t')}, we get

\vspace{-0.5cm}

\begin{equation}
    \dfrac{1}{3} \text{tr}\Bigl[\textbf{B}_{dis}(t+\Delta t/t^{'})\Bigl] = \dfrac{1}{3} \text{tr}\Bigl[J^{-2/3}(t+\Delta t/t)\textbf{F}(t+\Delta t/t) \textbf{B}_{dis}(t/t^{'}) \textbf{F}^{T}(t+\Delta t)\Bigr]. 
    \label{Simplification of 1/3 tr(Bdis(t+delta t/t)), step 1}
\end{equation}

\eqref{Simplification of 1/3 tr(Bdis(t+delta t/t)), step 1} can be rewritten using the identity tr(\textbf{M}\textbf{N}) = tr(\textbf{N}\textbf{M}) as 

\vspace{-0.5cm}

\begin{equation}
\begin{split}
    \dfrac{1}{3} \text{tr}\Bigl[\textbf{B}_{dis}(t+\Delta t/t^{'})\Bigr] &= \dfrac{1}{3} \text{tr} \Bigl[\textbf{B}_{dis}(t/t^{'}) \textbf{C}_{dis}(t+\Delta t/t)\Bigr].
    \label{Simplification of 1/3 tr(Bdis(t+delta t/t)), step 2}
\end{split}
\end{equation}

Again, for $\Delta t \rightarrow$ 0, using \eqref{Distortional part of C, identity for small delta t}, \eqref{Simplification of 1/3 tr(Bdis(t+delta t/t)), step 2} can be approximated as

\begin{equation}
    \dfrac{1}{3} \text{tr}\Bigl[\textbf{B}_{dis}(t+\Delta t/t^{'})\Bigr] \rightarrow \dfrac{1}{3} \text{tr}\Bigl[\textbf{B}_{dis}(t/t^{'})\Bigr].
    \label{Simplification of 1/3 tr(Bdis(t+delta t/t)), final step}
\end{equation}

Finally, for $\Delta t \rightarrow$ 0, substituting \eqref{Decomposition of Bdis(t+delta t/t')}, \eqref{Simplification of 1/3 tr(Bdis(t+delta t/t)), final step} in \eqref{Deviatoric part of Bdis(t+delta t/t')} and using \eqref{Distortional part of C, identity for small delta t}, we get the approximation

\vspace{-0.5cm}

\begin{equation}
\begin{split}
    \Bigl[\textbf{B}_{dis}(t+\Delta t/t^{'})\Bigr]_{0} & \rightarrow J^{-2/3}(t+\Delta t/t)\textbf{F}(t+\Delta t/t) \Bigl[\textbf{B}_{dis}(t/t^{'})\Bigr]_{0} \textbf{F}^{T}(t+\Delta t/t).
\end{split}
\label{Decomposition of deviatoric part of Bdis(t+delta t/t), final}
\end{equation}

Denoting the second term in \eqref{TB at t+delta t} as $\textbf{T}_2^{(B1)}$, we have

\begin{equation}
    \textbf{T}_2^{(B1)}(t+\Delta t) = \int_{0}^{t+\Delta t} \dfrac{n^{bd}_{ns}}{n^{bd}_s} k^{bd}_s e^{-k^{bd}_{ns} (t+\Delta t-t^{'})} \: \mu^{bd} \: \boldsymbol{\Gamma}^{bd}(t+\Delta t/t^{'}) \: dt^{'}.
    \label{TbB_2 at t+delta t, step 1}
\end{equation}

Equation \eqref{TbB_2 at t+delta t, step 1} can be decomposed into two integrals from (0 to $t$) and ($t$ to $t$ $+ \Delta t$) as

\vspace{-0.5cm}

\begin{equation}
\begin{split}
    \textbf{T}_2^{(B1)}(t+\Delta t) &= \int_{0}^{t} \dfrac{n^{bd}_{ns}}{n^{bd}_s} k^{bd}_s e^{-k^{bd}_{ns} (t+\Delta t-t^{'})} \: \mu^{bd} \: \boldsymbol{\Gamma}^{bd}(t+\Delta t/t^{'}) \: dt^{'} \\ 
    & + \int_{t}^{t+\Delta t} \dfrac{n^{bd}_{ns}}{n^{bd}_s} k^{bd}_s e^{-k^{bd}_{ns} (t+\Delta t-t^{'})} \: \mu^{bd} \: \boldsymbol{\Gamma}^{bd}(t+\Delta t/t^{'}) \: dt^{'}.   
\end{split}
\label{TbB_2 at t+delta t, step 2}
\end{equation}

Focusing on $\boldsymbol{\Gamma}^{bd}(t+\Delta t/t^{'})$, from \eqref{TB common terms convention definition}, we have

\vspace{-0.5cm}

\begin{equation}
    \boldsymbol{\Gamma}^{bd}(t+\Delta t/t^{'}) = J^{-1}(t+\Delta t/t^{'}) \dfrac{\overline{\lambda}^{bd}_L}{\overline{\lambda}(t+\Delta t/t^{'})} \mathcal{L}^{-1}\Bigl(\dfrac{\overline{\lambda}(t+\Delta t/t^{'})}{\overline{\lambda}^{bd}_L}\Bigr) \Bigl[\mathbf{B}_{dis}(t+\Delta t/t^{'})\Bigr]_0.
    \label{TB common terms convention definition (t+delta t/t'), step 1}
\end{equation}

For $\Delta t \rightarrow$ 0, using \eqref{Decomposition of F(t+delta t/t')}, \eqref{lambda(t+delta t/t') = lambda(t/t') for small delta t} and \eqref{Decomposition of deviatoric part of Bdis(t+delta t/t), final}, \eqref{TB common terms convention definition (t+delta t/t'), step 1} can be approximated as

\begin{equation}
    \boldsymbol{\Gamma}^{bd}(t+\Delta t/t^{'}) \rightarrow J^{-5/3}(t+\Delta t/t) \textbf{F}(t+\Delta t/t) \boldsymbol{\Gamma}^{bd}(t/t^{'}) \textbf{F}^{T}(t+\Delta t/t).
    \label{TB common terms convention definition (t+delta t/t'), step 2}
\end{equation}

Substituting \eqref{TB common terms convention definition (t+delta t/t'), step 2} in \eqref{TbB_2 at t+delta t, step 2}, we get the approximation

\vspace{-0.5cm}

\begin{equation}
\begin{split}
    \textbf{T}_2^{(B1)}(t+\Delta t) &\rightarrow e^{-k^{bd}_{ns} \Delta t} J^{-5/3}(t+\Delta t/t) \: \textbf{F}(t+\Delta t/t) \: \textbf{T}_2^{(B1)}(t) \: \textbf{F}^{T}(t+\Delta t/t) \\ 
    & + \int_{t}^{t+\Delta t} \dfrac{n^{bd}_{ns}}{n^{bd}_s} k^{bd}_s e^{-k^{bd}_{ns} (t+\Delta t-t^{'})} \: \mu^{bd} \: \boldsymbol{\Gamma}^{bd}(t+\Delta t/t^{'}) \: dt^{'}.   
\end{split}
\label{TbB_2 at t+delta t, step 3}
\end{equation}

Evaluating the second integral in \eqref{TbB_2 at t+delta t, step 3} using the trapezoidal rule and noting that $\boldsymbol{\Gamma}(t+\Delta t/t+\Delta t) = \textbf{0}$, \eqref{TbB_2 at t+delta t, step 3} can be rewritten as 

\vspace{-0.5cm}

\begin{equation}
\begin{split}
    \textbf{T}_2^{(B1)}(t+\Delta t) &\rightarrow e^{-k^{bd}_{ns} \Delta t} J^{-5/3}(t+\Delta t/t) \: \textbf{F}(t+\Delta t/t) \: \textbf{T}_2^{(B1)}(t) \: \textbf{F}^{T}(t+\Delta t/t) \\ 
    & + \dfrac{1}{2} \dfrac{n^{bd}_{ns}}{n^{bd}_s} k^{bd}_s e^{-k^{bd}_{ns} \Delta t} \: \mu^{bd} \: \boldsymbol{\Gamma}^{bd}(t+\Delta t/t) \: \Delta t.   
\end{split}
\label{TbB_2 at t+delta t, final step}
\end{equation}

Finally, the update procedure for $\textbf{T}^{(B1)}(t+\Delta t)$ under the limit $\Delta t \rightarrow$ 0  by combining \eqref{TB at t+delta t} and \eqref{TbB_2 at t+delta t, final step} can be summarized as

\vspace{-0.5cm}

\begin{equation}
\begin{split}
    \textbf{T}^{(B1)}(t+\Delta t) &\rightarrow e^{-k^{bd}_{ns} (t + \Delta t)} \mu^{bd} \boldsymbol{\Gamma}^{bd}(t+\Delta t/0) + \textbf{T}_{2}^{(B1)}(t+\Delta t).
\end{split}
\label{TB update procedure, final}
\end{equation}

Noting from \eqref{Contribution to T(B), B:O subchains (Mechanism B)} and \eqref{Contribution to T(B), physical subchains (Mechanism B)} that $\textbf{T}^{(B2)}(t+\Delta t)$ can be obtained by replacing the superscript ``$bd$" with ``$el$" in \eqref{TB at t+delta t}, the corresponding update procedure under the limit $\Delta t \rightarrow$ 0 following \eqref{TbB_2 at t+delta t, final step} and \eqref{TB update procedure, final} is given as

\begin{equation}
\begin{split}
    \textbf{T}^{(B2)}(t+\Delta t) &\rightarrow e^{-k^{el}_{ns} (t + \Delta t)} \mu^{el} \boldsymbol{\Gamma}^{el}(t+\Delta t/0) + \textbf{T}_{2}^{(B2)}(t+\Delta t),
\end{split}
\label{TB update procedure, final 2}
\end{equation}

with 

\begin{equation}
\begin{split}
    \textbf{T}_2^{(B2)}(t+\Delta t) &\rightarrow e^{-k^{el}_{ns} \Delta t} J^{-5/3}(t+\Delta t/t) \: \textbf{F}(t+\Delta t/t) \: \textbf{T}_2^{(B2)}(t) \: \textbf{F}^{T}(t+\Delta t/t) \\ 
    & + \dfrac{1}{2} \dfrac{n^{el}_{ns}}{n^{el}_s} k^{el}_s e^{-k^{el}_{ns} \Delta t} \: \mu^{el} \: \boldsymbol{\Gamma}^{el}(t+\Delta t/t) \: \Delta t.   
\end{split}
\label{TbB_2 at t+delta t, final step 2}
\end{equation}

\end{appendices}

\section*{Declaration of Interest}
The authors declare that they have no known competing financial interests or personal relationships that could have appeared to influence the work reported in this paper.

\newpage

\end{document}